\documentclass[11pt,letterpaper]{article}
\pdfoutput=1
\usepackage{jheppub}
\usepackage{color}
\usepackage{graphicx}

\usepackage{verbatim}
\usepackage{amsmath}
\usepackage{amssymb}
\usepackage{subfig}
\usepackage{url}
\usepackage{bbold}
\usepackage{slashed}

\usepackage{multirow}
\usepackage{threeparttable}
\usepackage{paralist}

\newcommand{\GeV}{\text{GeV}}

\newcommand{\ev}{\text{event}}
\newcommand{\jet}{\text{jet}}
\newcommand{\jets}{\text{jets}}
\newcommand{\subj}{\text{subjet}}
\newcommand{\subjs}{\text{subjets}}
\newcommand{\cut}{\text{cut}}
\newcommand{\trim}{\text{trim}}
\newcommand{\ptc}{p_{T{\rm cut}}}
\newcommand{\ptsubc}{p_{T{\rm subcut}}}

\newcommand{\sub}{\text{sub}}

\newcommand{\pythia}{\textsc{Pythia}}
\newcommand{\FastJet}{\textsc{FastJet}}
\newcommand{\MadGraph}{\textsc{MadGraph}}

\DeclareRobustCommand{\Sec}[1]{Sec.~\ref{#1}}

\DeclareRobustCommand{\App}[1]{App.~\ref{#1}}

\DeclareRobustCommand{\Fig}[1]{Fig.~\ref{#1}}

\DeclareRobustCommand{\Figss}[3]{Figs.~\ref{#1}, \ref{#2}, and \ref{#3}}
\DeclareRobustCommand{\Eq}[1]{Eq.~(\ref{#1})}
\DeclareRobustCommand{\Eqs}[2]{Eqs.~(\ref{#1}) and (\ref{#2})}

\DeclareRobustCommand{\Ref}[1]{Ref.~\cite{#1}}

\newcommand{\be}{\begin{equation}}
\newcommand{\ee}{\end{equation}}

\usepackage{xspace}

\newcommand{\beq}{\begin{eqnarray}}
\newcommand{\eeq}{\end{eqnarray}}

\newcommand{\F}{\mathcal{F}}
\newcommand{\Ft}{\widetilde{\mathcal{F}}}
\newcommand{\G}{\mathcal{G}}
\newcommand{\Gt}{\widetilde{\mathcal{G}}}
\newcommand{\HH}{\mathcal{H}}

\begin{document}

\title{Jet Observables Without Jet Algorithms}
\author{Daniele Bertolini,}
\author[*]{Tucker Chan,\note[*]{Deceased.}}
\author{and Jesse Thaler}
\affiliation{Center for Theoretical Physics, Massachusetts Institute of Technology, Cambridge, MA 02139, U.S.A.}
\emailAdd{danbert@mit.edu}
\emailAdd{jthaler@mit.edu}

\date{\today}

\abstract{We introduce a new class of event shapes to characterize the jet-like structure of an event.  Like traditional event shapes, our observables are infrared/collinear safe and involve a sum over all hadrons in an event, but like a jet clustering algorithm, they incorporate a jet radius parameter and a transverse momentum cut.  Three of the ubiquitous jet-based observables---jet multiplicity, summed scalar transverse momentum, and missing transverse momentum---have event shape counterparts that are closely correlated with their jet-based cousins.  Due to their ``local'' computational structure, these jet-like event shapes could potentially be used for trigger-level event selection at the LHC.  Intriguingly, the jet multiplicity event shape typically takes on non-integer values, highlighting the inherent ambiguity in defining jets.  By inverting jet multiplicity, we show how to characterize the transverse momentum of the $n$-th hardest jet without actually finding the constituents of that jet.  Since many physics applications do require knowledge about the jet constituents, we also build a hybrid event shape that incorporates (local) jet clustering information.  As a straightforward application of our general technique, we derive an event-shape version of jet trimming, allowing event-wide jet grooming without explicit jet identification.  Finally, we briefly mention possible applications of our method for jet substructure studies.}

%\keywords{keyword one, keyword two}
%\arxivnumber{XXXX.XXXX}
\preprint{MIT-CTP {4502}}

\maketitle

\section{Introduction}
\label{sec:introduction}

When quarks and gluons are produced in high energy particle collisions, they undergo a process of showering and hadronization, and the resulting final state can be organized in terms of clusters of hadrons called jets. Jets play a key role at experiments like the Large Hadron Collider (LHC), both for testing standard model (SM) physics and for searching for new phenomena beyond the SM.   At present, most jet studies at the LHC are based on jets identified with a jet algorithm \cite{Ellis:2007ib,Salam:2009jx}.  Algorithms such as anti-k$_T$ \cite{Cacciari:2008gp} cluster final state hadrons into jet objects, whose four-momenta are then used as inputs for subsequent analyses.  An alternative approach is provided by event shape observables, which are functions involving all final state hadrons in a collision event.  Event shapes were extensively used for precision tests of quantum chromodynamics (QCD) at $e^+ e^-$ colliders \cite{Dasgupta:2003iq,Heister:2003aj,Abdallah:2003xz,Achard:2004sv,Abbiendi:2004qz}, and various event shapes have been proposed and used at hadron colliders \cite{Banfi:2004nk,Banfi:2010xy,Aaltonen:2011et,Khachatryan:2011dx,Aad:2012np}.

In this paper, we will blur the distinction between jet algorithms and event shapes by constructing jet-like event shapes.  These event shapes incorporate a jet-like radius $R$ as well as a jet-like transverse momentum cut $\ptc$, and they can be viewed as counterparts to some of the most commonly used jet-based observables.  While these event shapes do not involve any kind of clustering procedure, they are correlated with their jet-based cousins and yield comparable information about the jet-like structure of an event.  In this paper, we will mainly discuss jet-like event shapes, but the generalization to subjet-like jet shapes is straightforward, with potential applications in jet substructure studies \cite{Abdesselam:2010pt,Altheimer:2012mn}.

We will start by constructing three jet-like event shapes that mirror the three inclusive jet observables---jet multiplicity, summed scalar transverse momentum, and missing transverse momentum---that appear ubiquitously in jet studies at both the trigger and analysis levels.  For example, we will construct the jet multiplicity event shape as 
\be
\label{eq:Njetintro}
\widetilde{N}_\jet(\ptc,R) = \sum_{i\in\ev}\frac{p_{T i}}{p_{T i,R}}\Theta(p_{T i,R}-\ptc),
\ee
where $p_{T i,R}$ is the transverse momentum contained in a cone of radius $R$ around particle $i$.  Our technique for building jet-like event shapes can be generalized to a broad class of inclusive jet observables, namely observables built as a sum over all jets in an event.

We will then show how to manipulate these event shapes to characterize individual jets.  By inverting $\widetilde{N}_\jet$, we can characterize the $p_T$ of the $n$-th hardest jet without explicitly identifying the set of hadrons that form that jet.  Of course, for practical jet studies, one often wants to know the actual constituents of a jet.  Since our jet-like event shapes do not have a natural clustering interpretation, we develop a hybrid method that incorporates local jet clustering into an ``event shape density''.   The integral over this density gives the corresponding event shape, but the density distribution itself has spikes in the direction of candidate jet axes.

A perhaps surprising application of our method is for jet grooming \cite{Butterworth:2008iy,Ellis:2009su,Ellis:2009me,Krohn:2009th}.  Jet grooming methods aim to mitigate the effects of jet contamination from initial state radiation, underlying event, and pileup by removing soft wide-angle radiation from a jet.  In the case of pileup, one can use jet grooming in concert with area subtraction techniques \cite{Cacciari:2007fd,Cacciari:2008gn,Soyez:2012hv}.  Here, we show how jet trimming \cite{Krohn:2009th} can be recast as an event shape.  Our method is equivalent to assigning a weight to every particle in the event of
\be
w_i = \Theta\left(\frac{p_{T i,R_\sub}}{p_{T i,R}}-f_\text{cut}\right) \Theta(p_{T i,R}-\ptc).
\ee
This ``shape trimming'' method involves the same $f_\cut$ and $R_\sub$ parameters as the original ``tree trimming'' procedure, but does not require the explicit identification of jets or subjets.

There are a number of potential applications for these jet-like event shapes.  At the trigger level, they offer a ``local'' way to characterize the gross properties of an event.  By local, we mean that the event shape is defined as a sum over regions of interest of radius $R$, without needing global clustering information.  This local structure allows for efficient parallel computation of the event shape.\footnote{We thank David Strom for pointing out this possibility to us.}  If desired, one could even include (local) pileup suppression at the trigger level by incorporating (local) trimming.   At the analysis level, these event shapes offer an alternative way to characterize jets in regions of phase space where jets are overlapping.  In particular, whereas standard jet algorithms always give an integer value for the jet multiplicity $N_\jet$, the corresponding event shape $\widetilde{N}_\jet$ in \Eq{eq:Njetintro} typically returns a non-integer value, reflecting the inherent ambiguity in defining jets.  At minimum, one can use these event shapes to test the robustness of standard jet selection criteria, since a cut on the jet-like event shape should give similar results to a cut on jet objects for the same value of $R$ and $\ptc$.  Ultimately, one would like to study the analytic properties of these jet-like event shapes in perturbative QCD, though such studies are beyond the scope of this paper.

It is worth noting that our approach shares some of the same goals and features as other jet-like methods.  For defining jet observables through event shapes, there has been previous work showing how to construct effective jet clustering procedures via optimization of event shapes \cite{Ellis:2001aa}, most recently in taking $N$-jettiness \cite{Stewart:2010tn} and minimizing over the choice of jet axes \cite{Thaler:2011gf}.  The difference here is that the jet-like event shapes do not have an obvious clustering interpretation.  There are also methods that cast jet finding as a more general optimization problem \cite{Berger:2002jt,Angelini:2002et,Angelini:2004ac,Grigoriev:2003yc,Grigoriev:2003tn,Chekanov:2005cq,Lai:2008zp,Volobouev:2009rv,Ellis:2012sn,Kahawala:2013sba}, often with a probabilistic interpretation of an event.  The difference here is that we (uniquely) assign an event shape value to each event.  A set of variables that avoids explicit jet clustering are energy correlation functions \cite{Larkoski:2013eya}, which can characterize an event's structure without reference to even a jet axis (in contrast to $N$-jettiness), though different  correlation functions are needed for different jet multiplicities.  The difference here is that we need not specify the jet multiplicity of interest, though we do need to choose the jet radius $R$ and threshold $\ptc$.  Finally, for giving a global characterization of an event, there has been recent work to describe the jet-like nature of an event by summing over the contributions of large radius jets \cite{Hook:2012fd,Cohen:2012yc,Hedri:2013pvl}, though these observables make explicit use of tree-like recursive jet algorithms. The difference here is that we can achieve a similar global characterization through an inclusive sum over all particles in an event.

The rest of this paper is organized as follows.  We define event shapes for inclusive jet observables in \Sec{sec:JetObs} and perform Monte Carlo studies to demonstrate the correlations present with their jet-based cousins.  We then show in \Sec{sec:indivjet} how to manipulate and modify these event shapes to characterize the properties of individual jets, in particular how to find the jet constituents using a hybrid event shape density with a ``winner-take-all'' recombination scheme.  We describe our shape trimming technique in \Sec{sec:Trimming} and show how it is closely correlated with ordinary tree trimming.  We suggest possible generalization of our method in \Sec{sec:Generalizations} and draw conclusions in \Sec{sec:Conclusions}.  All of the event shapes described in this paper are available as an add-on to \FastJet~3  \cite{Cacciari:2011ma} as part of the \FastJet~contrib project (\url{http://fastjet.hepforge.org/contrib/}).

\section{Event Shapes for Inclusive Jet Observables}
\label{sec:JetObs}

Jet multiplicity ($N_\jet$), summed scalar transverse momentum ($H_T$), and missing transverse momentum ($\slashed{p}_T$) are three of the most ubiquitous observables used to globally characterize an event with jets in the final state.  Given jets identified through some jet algorithm with characteristic radius $R$, they are defined as
\begin{align}
N_\jet(\ptc,R) &= \sum_\jets \Theta(p_{T\jet}-\ptc), \label{eq:Njet} \\
H_T(\ptc,R) &= \sum_\jets p_{T \jet}\, \Theta(p_{T\jet}-\ptc), \\
\slashed{p}_T(\ptc,R) &=  \left| \sum_\jets \vec{p}_{T \jet} \, \Theta(p_{T \jet}-\ptc) \right|, \label{eq:pTmiss}
\end{align}
where $\vec{p}_{T \jet}$ is the transverse momentum measured with respect to the beam axis, $p_{T \jet} = |\vec{p}_{T \jet}|$, and $\ptc$ is the $p_T$ threshold for the analysis.\footnote{Typically, $\slashed{p}_T$ would include non-hadronic objects in the event as well, but we will not need that for the case studies in this paper.}  We have made the arguments $\ptc$ and $R$ explicit in anticipation of the discussion in \Sec{sec:indivjet}.  These three observables are part of a broader class of inclusive jet observables
\be
\label{eq:defF}
\F(\ptc,R) = \sum_\jets \F_\jet \,  \Theta(p_{T \jet}-\ptc),
\ee
where $\F_\jet=f(\{p^\mu_j\}_{j\in\jet})$ depends on the kinematics of the individual jet constituents.

\begin{figure}[t]
  \centering
    \includegraphics[scale=0.55]{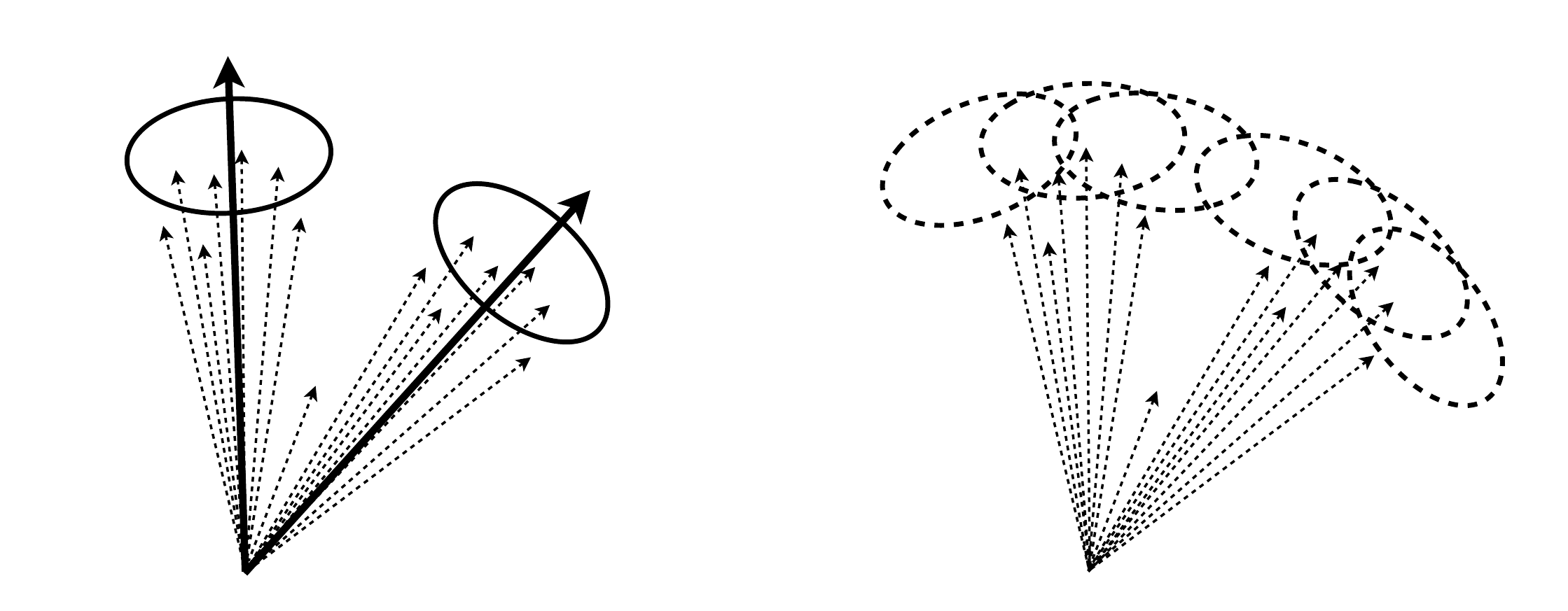}
  \caption{Instead of defining inclusive jet observables by summing over jet regions according to a jet algorithm (left), our event shapes sum over the contributions from cones of radius $R$ centered on each particle $i$ (right). The weight factor $p_{Ti}/p_{Ti,R}$ in \Eq{eq:generalForm} avoids double-counting despite overlapping cones.  For infinitely narrow jets separated by more than $R$, the two methods yield the same result.}
  \label{fig:strategy}
 \end{figure}

As written, $\F$ is intrinsically tied to a given jet algorithm.  Here, we wish to build a corresponding event shape $\Ft$ which makes no reference to a clustering procedure.  The first step is to effectively replace the sum over jets with a sum over particles, using the fact that
\be
\label{eq:repl1}
1 = \frac{1}{p_{T \jet}} \sum_{i \in \jet} p_{T i}, \qquad  \sum_\jets \sum_{i \in \jet} \Rightarrow \sum_{i \in \ev},
\ee
where we now use a more convenient definition $p_{T \jet} \equiv \sum_{i \in \jet} p_{T i}$ such that the first expression is a strict equality,\footnote{Note that the two definitions $p_{T \jet} \equiv |\vec{p}_{T \jet}|$ vs.\ $\sum_{i \in \jet} p_{T i}$ yield the same value for infinitely narrow jets.  Instead of $p_{T}$, one could accomplish the same goal using the energy relation $1 = (1/E_\jet) \sum_{i \in \jet} E_{i}$.} and the second expression has an implicit restriction to particles $i$ which are part of a jet cluster.  The second step is to convert jet measurements into measurements on jet-like cones of radius $R$ around each particle:
\begin{align}
\label{eq:repl2}
\F_\jet=f(\{p^\mu_j\}_{j\in\jet}) \quad &\Rightarrow \quad \F_{i,R} = f(\{p^\mu_j\,\Theta(R-\Delta R_{ij})\}_{j\in\ev}), \\
\label{eq:pTR}
p_{T\jet} = \sum_{i \in \jet} p_{T i} \quad &\Rightarrow \quad p_{T i,R} = \sum_{j\in\ev}p_{T j} \, \Theta(R-\Delta R_{ij}),
\end{align}
where $\Delta R_{ij}=\sqrt{\Delta\eta_{ij}^2+\Delta\phi_{ij}^2}$ is the distance in the rapidity-azimuth plane, and $p_{T i,R}$ is the sum of transverse momentum contained in a cone of radius $R$ around particle $i$. 
Applying these two steps, we derive the event shape associated with the generic inclusive jet observable in \Eq{eq:defF}:
\be
\label{eq:generalForm}
\Ft(\ptc,R)=\sum_{i\in\ev}\frac{p_{Ti}}{p_{Ti,R}}\F_{i,R}\,\Theta(p_{T i,R}-\ptc).
\ee
Because of the weight factor $p_{Ti}/p_{Ti,R}$, this definition avoids double-counting, even though the jet-like cones around each particle are overlapping.  As long as the original $\F_\jet$ was infrared/collinear safe, then $\Ft$ will also be infrared/collinear safe (assuming $\ptc > 0$).  Our general strategy is depicted in \Fig{fig:strategy}.

Following this logic, we define the following jet-like event shapes corresponding to $N_\jet$, $H_T$, and $\slashed{p}_T$:
\begin{align}
\widetilde{N}_\jet(\ptc,R) & = \sum_{i\in\ev}\frac{p_{T i}}{p_{T i,R}}\Theta(p_{T i,R}-\ptc), \\
\widetilde{H}_T(\ptc,R) & = \sum_{i\in\ev} p_{T i}\, \Theta(p_{T i,R}-\ptc),\\
\widetilde{\slashed{p}}_T(\ptc,R) & = \left|\sum_{i\in\ev} \vec{p}_{T i}\, \Theta(p_{T i,R}-\ptc)\right|,\label{eq:ptMiss}
\end{align}
where $p_{T i,R}$ is defined in \Eq{eq:pTR}.  For the sake of simplicity, in \Eq{eq:ptMiss} we approximated $\vec{p}_{Ti,R}\approx p_{Ti,R}\,\hat{p}_{Ti}$, which is strictly true only for infinitely narrow jets.\footnote{Alternatively, one could recover \Eq{eq:ptMiss} by noticing that if we assume $\F_\jet\equiv\vec{p}_{T\jet}\simeq\sum_{j\in\jet}\vec{p}_{Tj}$, then we can skip the first replacement in \Eq{eq:repl1}, and directly convert the double sum into a sum over the event.}
For events consisting of infinitely narrow jets separated by more than $R$, the event shapes $\widetilde{N}_\jet$, $\widetilde{H}_T$, and $\widetilde{\slashed{p}}_T$ yield identical values to their jet-based counterparts $N_\jet$, $H_T$, and $\slashed{p}_T$.  We describe applications and generalizations of this procedure to other inclusive jet (and subjet) observables in \Sec{sec:Generalizations}.

To get a sense for how these event shapes behave, it is useful to study how they correlate with their jet-based counterparts.  For this study, we generate event samples for the $\sqrt{s}=8$~TeV LHC in \MadGraph~5 \cite{Alwall:2011uj}, with showering and hadronization carried out in \pythia~8.157 \cite{Sjostrand:2007gs}.\footnote{Unless otherwise specified, this will be the standard setup for Monte Carlo studies throughout the paper.}  For the standard jet-based observables, we use \FastJet~3.0.2 with the anti-$k_T$ jet algorithm \cite{Cacciari:2008gp} with a jet radius $R = 0.6$ and $\ptc = 25~\GeV$. For the event shapes, we use the same value of $R$ and $\ptc$. In order to (artificially) highlight the behavior of our event shapes on both one jet and two jet events, we set the minimum $p_T$ at the parton level in \MadGraph\ to $p_{T\text{cut}}^\text{parton}=25~\GeV$.\footnote{Without a $p_{T\text{cut}}^\text{parton}$ cut, there would of course be more one jet than two jet events.  We checked that the event shape distributions remain correlated with their jet-based counterparts as $p_{T\text{cut}}^\text{parton} \to 0$.}

\begin{figure}[tp]
  \centering
  \subfloat[]{\label{fig:HistoNj}
    \includegraphics[scale=0.65]{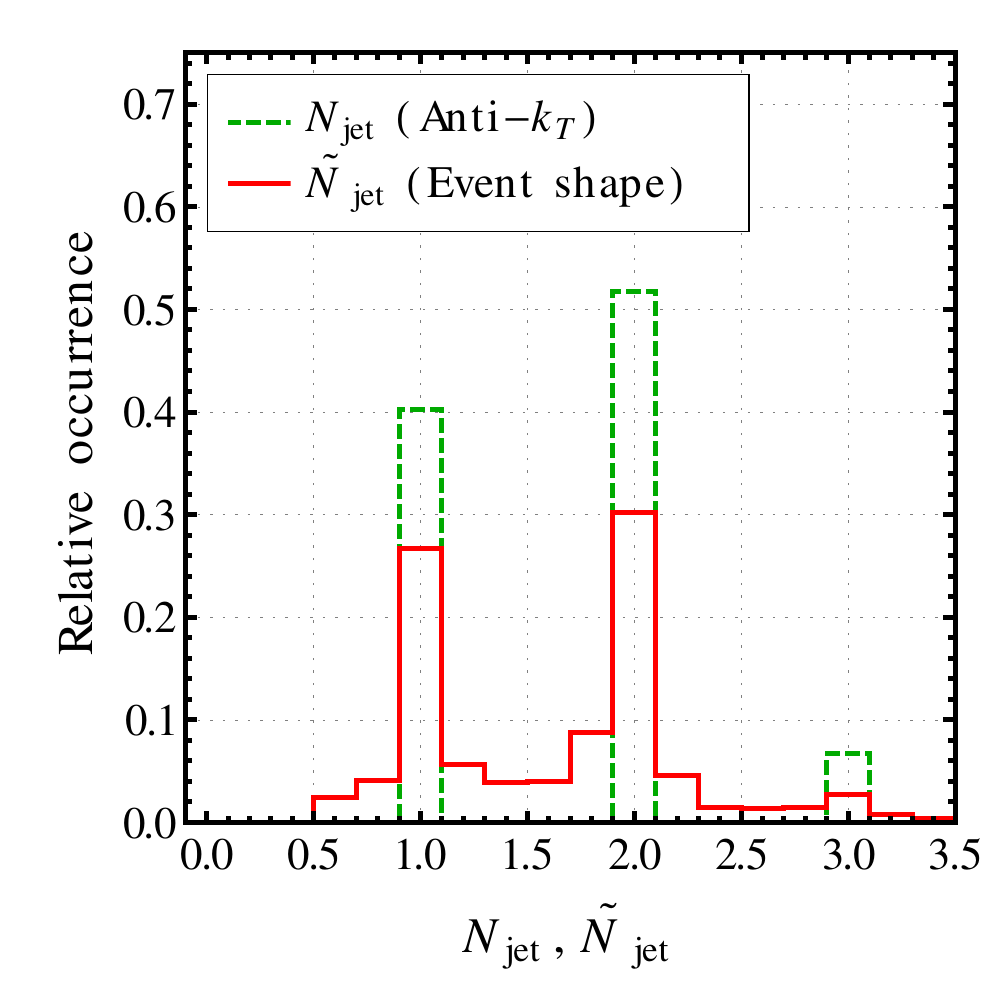}}   
\hspace{0.3in}
  \subfloat[]{\label{fig:ScatterNj}
    \includegraphics[scale=0.65]{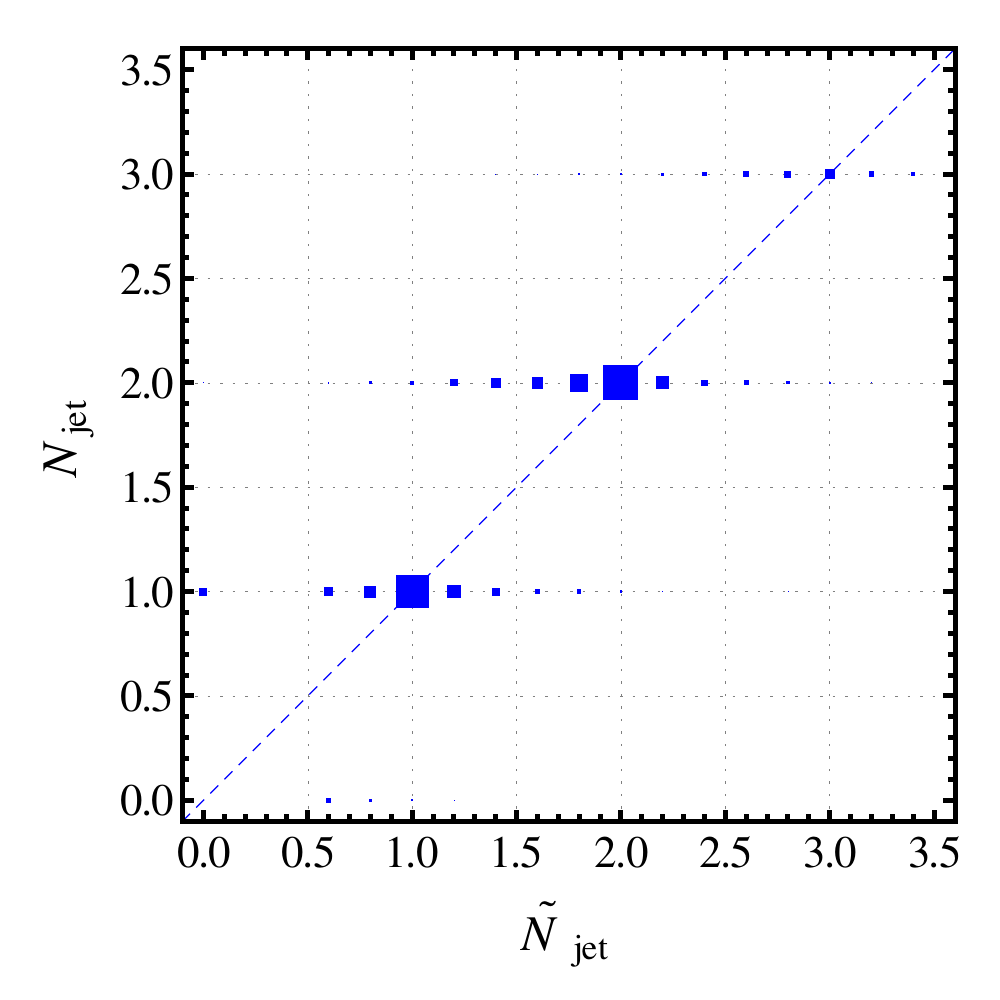}}
  \caption{Jet multiplicity (i.e.\ $N_{\jet}$) for QCD dijet events. \Fig{fig:HistoNj} shows the distribution of the number of anti-$k_T$ jets with $R=0.6$ and  $\ptc=25$ GeV (green dashed  curve), and of the corresponding event shape with the same values of $R$ and $\ptc$ (red curve).  Only events with $N_\jet\geq 1$ or $\widetilde{N}_\jet\geq 0.5$ are shown, and a parton level cut of $p_{T\text{cut}}^\text{parton}=25~\GeV$ is employed to give a reasonable sample of both one jet and two jet events. Whereas $N_{\jet}$ takes on only integer values, the event shape $\widetilde{N}_{\jet}$ is continuous, albeit with spikes near integer values. \Fig{fig:ScatterNj} shows the correlation between the two observables, where the area of the squares is proportional to the fraction of events in each bin. In the correlation plot, events that fail one of the jet cut criteria are assigned the corresponding value of zero.}
  \label{fig:Njet}
 \end{figure}

\begin{figure}[tp]
  \centering
  \subfloat[]{\label{fig:HistoHT}
    \includegraphics[scale=0.65]{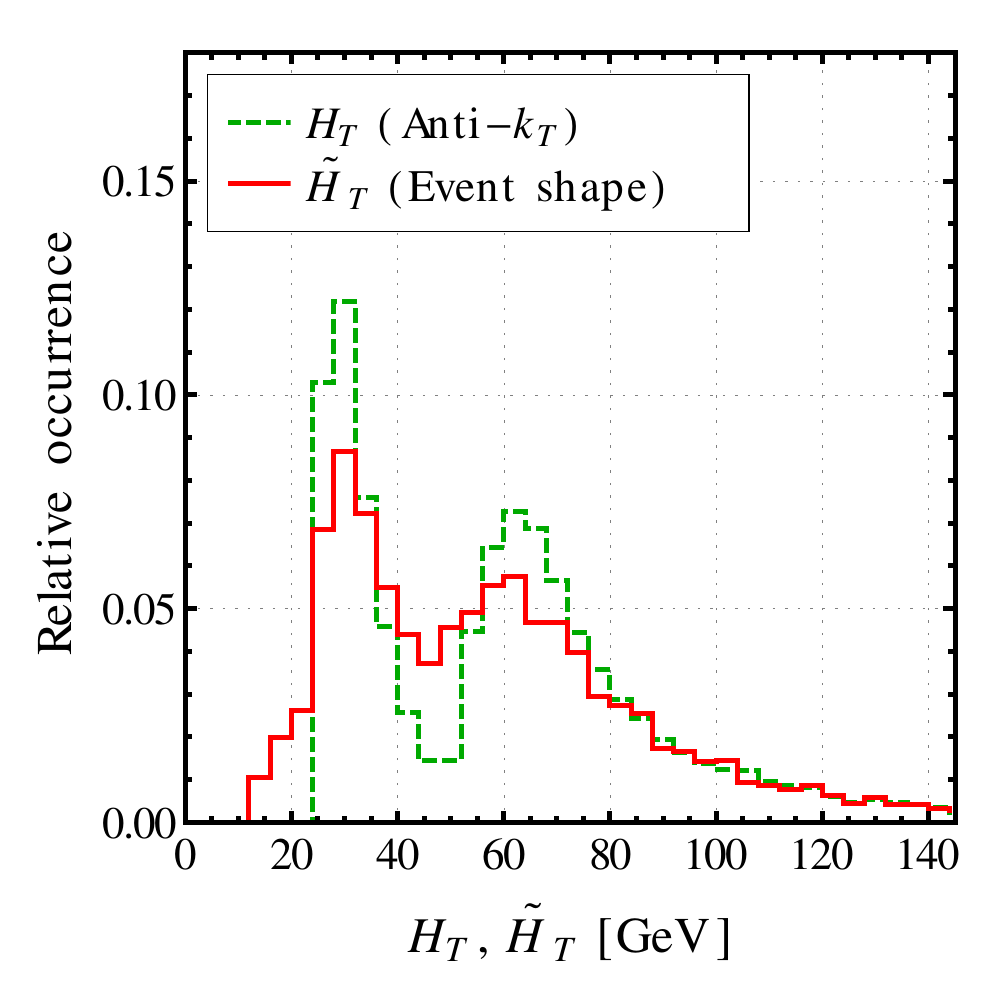}}   
\hspace{0.3in}
  \subfloat[]{\label{fig:ScatterHT}
    \includegraphics[scale=0.65]{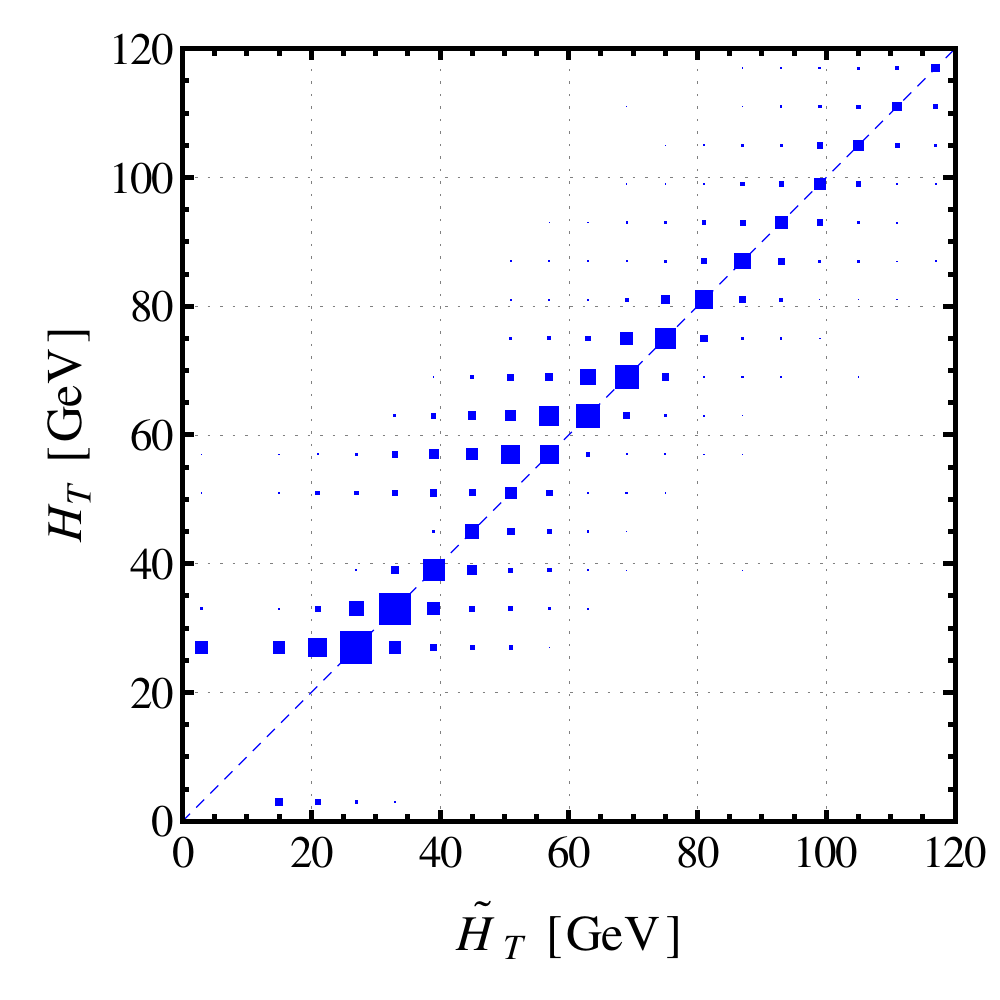}}
  \caption{Summed scalar transverse momentum (i.e.\ $H_T$) for QCD dijet events.  The jet parameters, formatting, and cuts are the same as for \Fig{fig:Njet}.  Because of the smoother behavior of the event shape $\widetilde{H}_T$, the peaks rising at $\ptc$ and $2\, \ptc$ are less pronounced than for $H_T$.}
  \label{fig:HT}
 \end{figure}

  \begin{figure}[tp]
  \centering
  \subfloat[]{\label{fig:HistoMET}
    \includegraphics[scale=0.65]{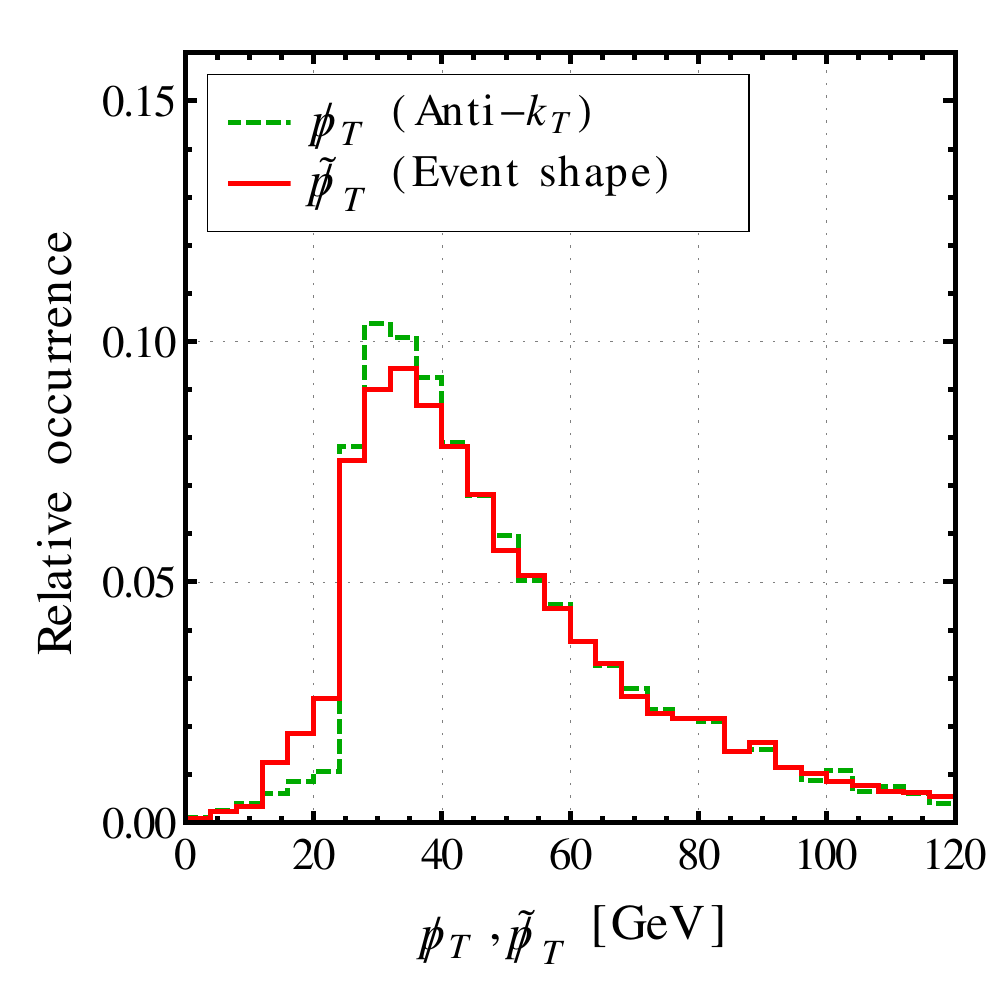}}   
\hspace{0.3in}
  \subfloat[]{\label{fig:ScatterMET}
    \includegraphics[scale=0.65]{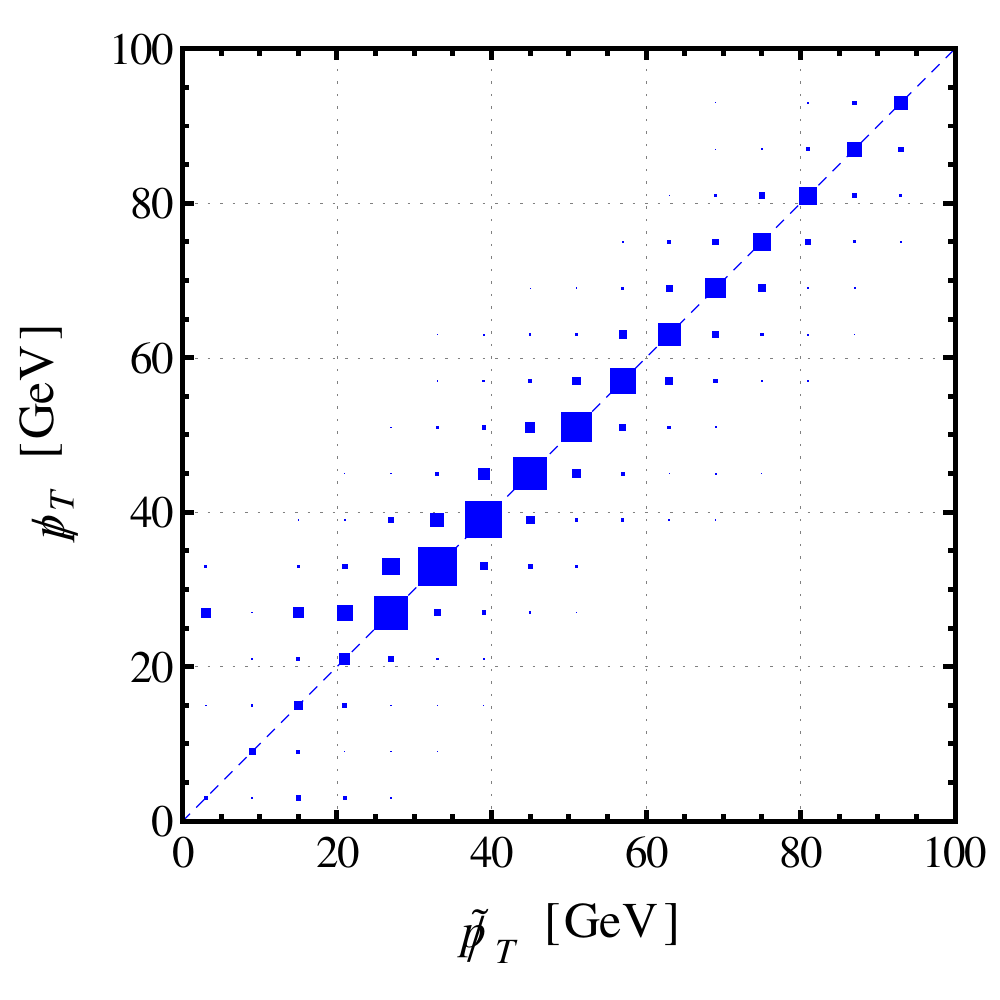}}
  \caption{Missing transverse momentum (i.e.\  $\slashed{p}_T$) for $Z(\rightarrow \nu\bar{\nu})+j$ events. The jet parameters, formatting, and cuts are the same as for \Fig{fig:Njet}. Again, we see a smoother turn on behavior for $\widetilde{\slashed{p}}_T$ compared to $\slashed{p}_T$.}
  \label{fig:pTmiss}
 \end{figure}

In \Fig{fig:Njet}, we compare $N_\jet$ versus $\widetilde{N}_\jet$ for QCD dijet events.  Whereas $N_\jet$ takes on discrete values, $\widetilde{N}_\jet$ yields a continuous distribution, though the observables are correlated on an event-by-event basis.  Here and in the following plots we only show events with $N_\jet\geq 1$ and $\widetilde{N}_\jet\geq 0.5$; the choice of the lower limit on $\widetilde{N}_\jet$ will be justified in \Sec{subsec:jetPt}.  In \Fig{fig:HT}, we compare $H_T$ versus $\widetilde{H}_T$ again for QCD dijet events.  Because of the $\ptc = 25~\GeV$ cut, $H_T$ exhibits two spikes that rise starting at $25~\GeV$ (for one jet events) and $50~\GeV$ (for two jet events), whereas $\widetilde{H}_T$ is smoother in this turn-on region.\footnote{With $p_{T\text{cut}}^\text{parton} \to 0$, the same features are visible, albeit with the one jet spike being much larger than the two jet spike.}   In the tail region, the distributions of $H_T$ and $\widetilde{H}_T$ are very similar. In \Fig{fig:pTmiss}, we compare $\slashed{p}_T$ versus $\widetilde{\slashed{p}}_T$ for $Z$ plus jet events where the $Z$ decays to neutrinos.  Again we see a spike that rises starting at $25~\GeV$ for $\slashed{p}_T$ which is milder in the event shape $\widetilde{\slashed{p}}_T$, though the distributions are quite similar throughout.

  \begin{figure}[tp]
  \centering
  \subfloat[]{\label{fig:HistoR}
    \includegraphics[scale=0.65]{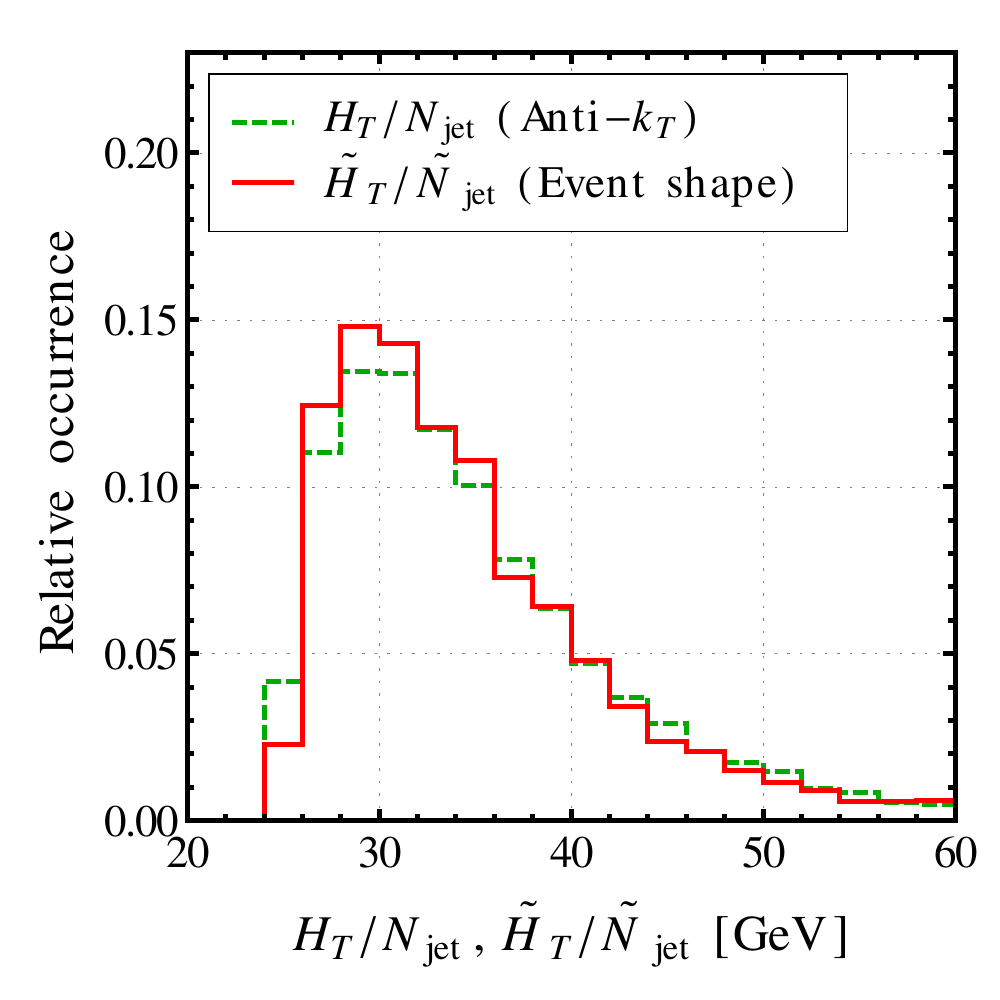}}   
\hspace{0.3in}
  \subfloat[]{\label{fig:ScatterR}
    \includegraphics[scale=0.65]{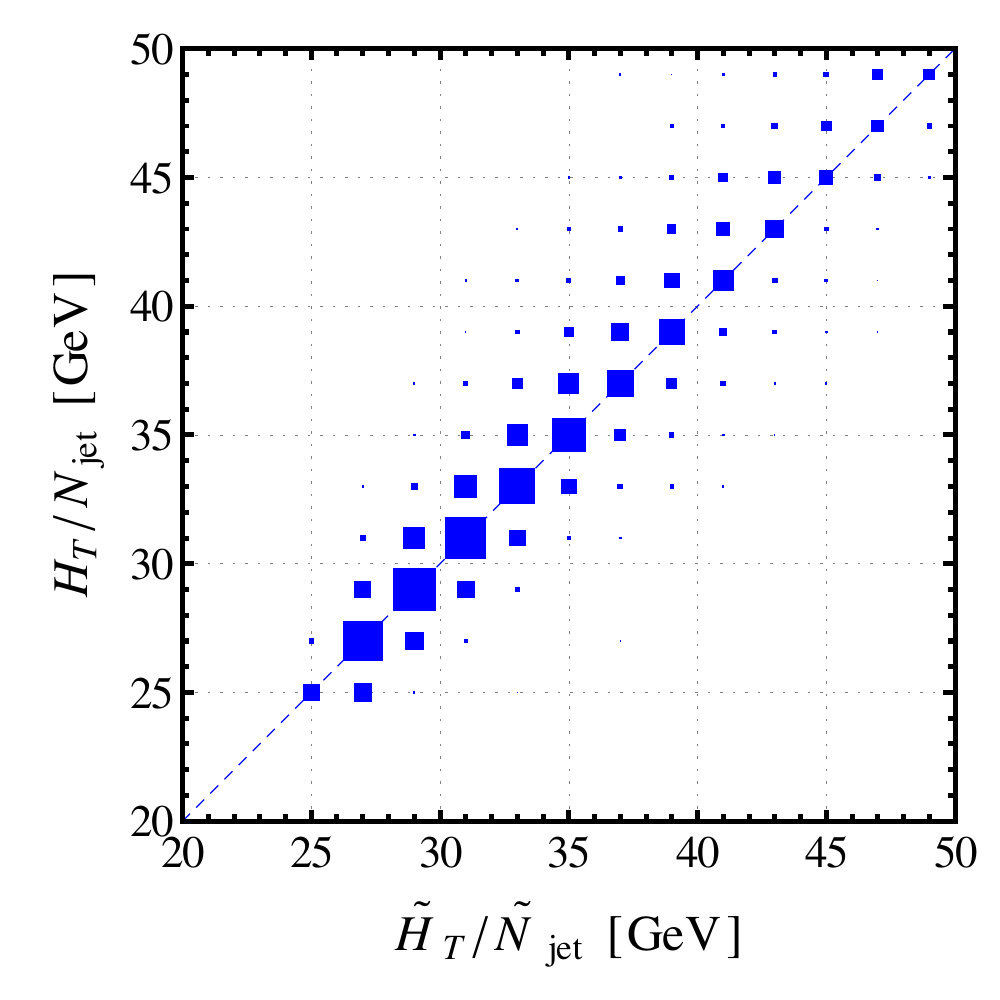}}
  \caption{Average jet transverse momentum (i.e.\ $H_T$ divided by $N_\jet$) for QCD dijet events.  The jet parameters, formatting, and cuts are the same as for \Fig{fig:Njet}. }
  \label{fig:HToverN}
 \end{figure}

 \begin{figure}[tp]
  \centering
  \subfloat[]{\label{fig:HistoMETSIG}
    \includegraphics[scale=0.65]{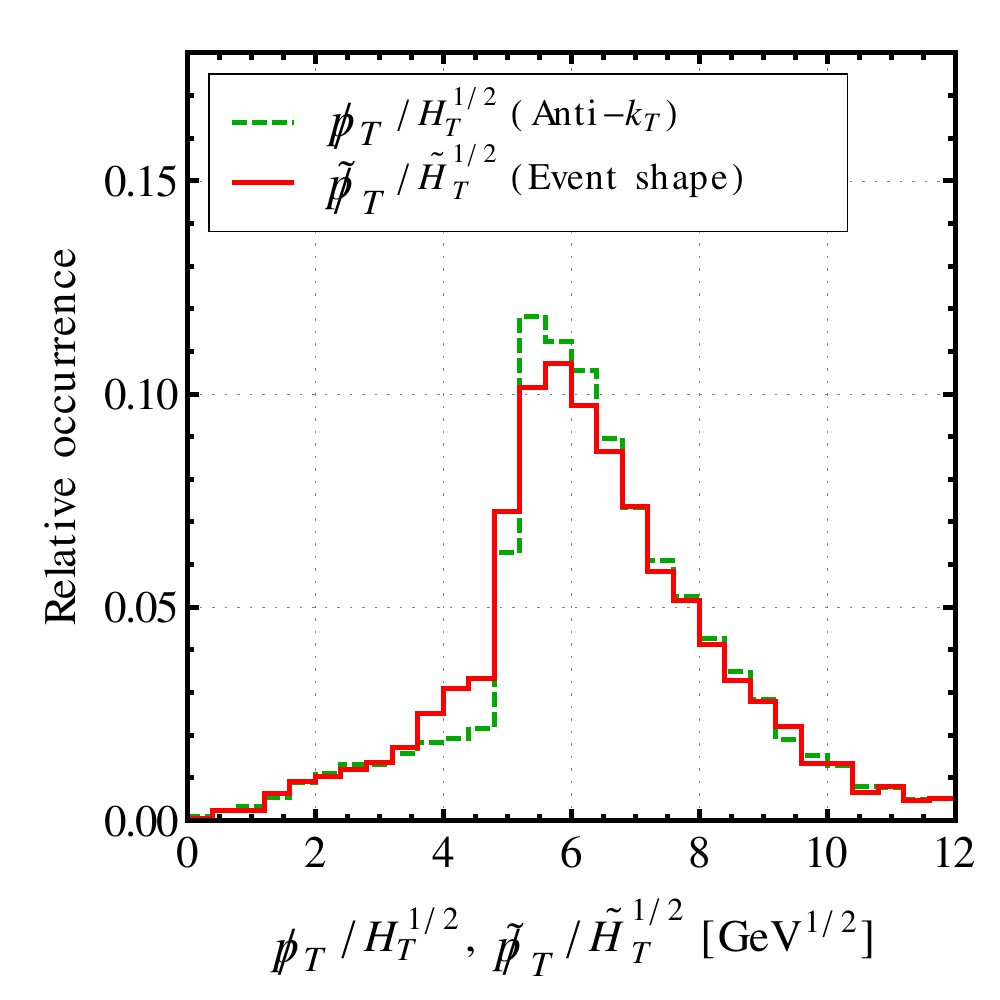}}   
\hspace{0.3in}
  \subfloat[]{\label{fig:ScatterMETSIG}
    \includegraphics[scale=0.65]{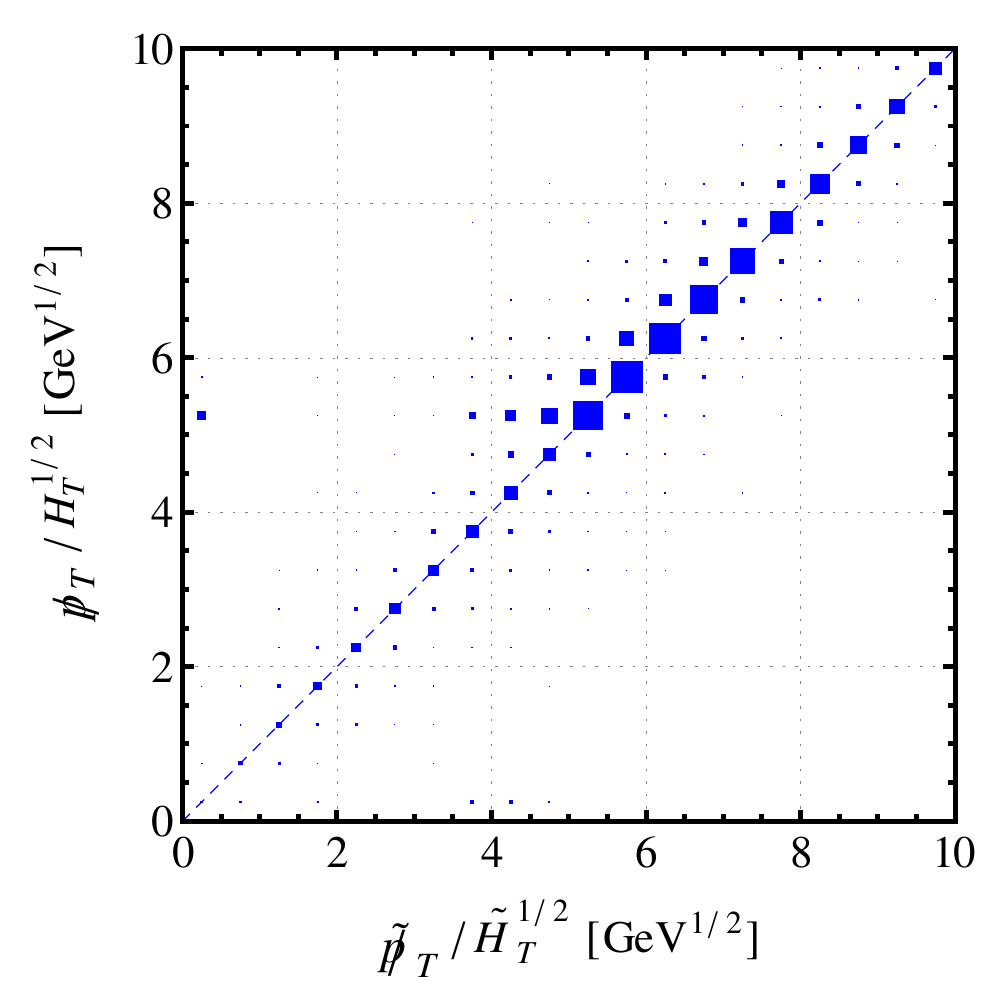}}
  \caption{Missing transverse momentum significance (i.e.\ $\slashed{p}_T$ divided by $\sqrt{H_T}$) for $Z(\rightarrow \nu\bar{\nu})+j$ events. The jet parameters, formatting, and cuts are the same as for \Fig{fig:Njet}. }
  \label{fig:METSIG}
 \end{figure}

Just as for ordinary jet-based observables, one can construct interesting composite functions with the event shapes.  For example, one can consider the average $p_T$ of the jets in an event, and we compare $H_T/N_\jet$ versus  $\widetilde{H}_T/\widetilde{N}_\jet$ in \Fig{fig:HToverN}.  Another useful composite variable is missing $p_T$ significance \cite{DZeroNote3629,Nachman:2013bia}, and we compare $\slashed{p}_T/\sqrt{H_T}$ versus $\widetilde{\slashed{p}}_T/\sqrt{\widetilde{H}_T}$ in \Fig{fig:METSIG}.

The differences between the jet-like event shapes and their jet-based counterparts reflects the intrinsic ambiguity in how to define a jet, seen most strikingly in the fact that $\widetilde{N}_\jet$ does not take on integer values.   For jet observables that are inclusive over all jets, $\widetilde{N}_\jet$, $\widetilde{H}_T$, and $\widetilde{\slashed{p}}_T$ characterize the global properties of the event without defining a clustering procedure, and appear to give very similar information to $N_\jet$, $H_T$, and $\slashed{p}_T$ for the same values of $R$ and $\ptc$.  Of course, because there is no clustering, one cannot determine the kinematics of any individual jet with the event shape alone (see however \Sec{sec:indivjet} below).
In terms of computational costs, the bottleneck is calculating $p_{T i,R}$ in \Eq{eq:pTR} for every particle $i$, which naively scales like $N^2$ for an event with $N$ hadrons, though the computational costs are dramatically reduced if one has an efficient way to determine which particles are within a radius $R$ of particle $i$.\footnote{In our \FastJet\ add-on, we make a crude attempt in this direction by partitioning the event into overlapping blocks of size $2R \times 2R$ and by caching the results of repeated calculations. Our implementation could potentially be further optimized by using, for example, an alternative distance heuristic.}  In practice, calculating $\widetilde{N}_\jet$ using our \FastJet~3 add-on with a standard laptop takes about as long as calculating $N_\jet$ with anti-$k_T$.  Moreover, $\widetilde{N}_\jet$ can be parallelized since it only depends on the contributions from particles within a radius $R$ (i.e. it is defined ``locally''). This feature makes it possible to implement $\widetilde{N}_\jet$ in a low-level trigger for sufficiently small $R$.  The key question at the trigger level is whether an event-shape-based trigger has better properties (e.g.~turn-on, stability, calibration, etc.)\ than a jet-based trigger, but a detailed study of this issue is beyond the scope of this work.

\section{Characterizing Individual Jets}
\label{sec:indivjet}

While inclusive jet observables are useful for characterizing the gross properties of an event, one would still like to gain more exclusive information about the kinematics of individual jets.  In general, our jet-like event shapes do not yield that kind of exclusive information, but we will demonstrate a novel way to extract the (approximate) transverse momentum of individual jets by using the full functional form of $\widetilde{N}_\jet$.  We will then define a hybrid event shape density that incorporates (local) jet clustering information in order to determine the constituents of individual jets.

\subsection{Jet Transverse Momentum}
\label{subsec:jetPt}
Consider the jet multiplicity event shape $\widetilde{N}_\jet(\ptc,R)$.  As shown in \App{app:invert}, there is a computationally efficient way to find the pseudo-inverse of this function with respect to $\ptc$, namely $\ptc(\widetilde{N}_\jet, R)$.\footnote{The reason this is a pseudo-inverse is that $\widetilde{N}_\jet(\ptc,R)$ is a monotonically decreasing step-wise function of $\ptc$, so there is a range of values of $\ptc$ with the same $\widetilde{N}_\jet$.  Once the values of $p_{T i,R}$ are known, the algorithm in \App{app:invert} scales like $N \log N$ for $N$ particles.}  We will see in a moment that it is useful to introduce an offset $n_{\rm off}$, so we define
\be
\label{eq:noffset}
\widetilde{p}_T(n,R) = \ptc(n - n_{\rm off}, R) \quad \text{with} \quad 0 \lesssim n_{\rm off} \lesssim 1,
\ee
where the default value of $n_{\rm off}$ is $0.5$.  The corresponding function for ordinary jets is denoted $p_T(n,R)$.

The function $\widetilde{p}_T(n,R)$ effectively gives the $p_T$ of the $n$-th hardest jet.  That is, it gives the value of the $p_T$ threshold needed to include the $n$-th jet's contribution to $\widetilde{N}_\jet$.  For infinitely narrow jets separated by more than $R$, $\ptc(\widetilde{N}_\jet, R)$ takes discrete jumps as $\widetilde{N}_\jet$ increases by integer values.  More generally, the offset $n_{\rm off}$ accounts for the fact that an event with $n$ jets most likely returns a value of $\widetilde{N}_\jet$ between $n-1$ and $n$.

 \begin{figure}[t]
  \centering
  \subfloat[]{
    \includegraphics[scale=0.50]{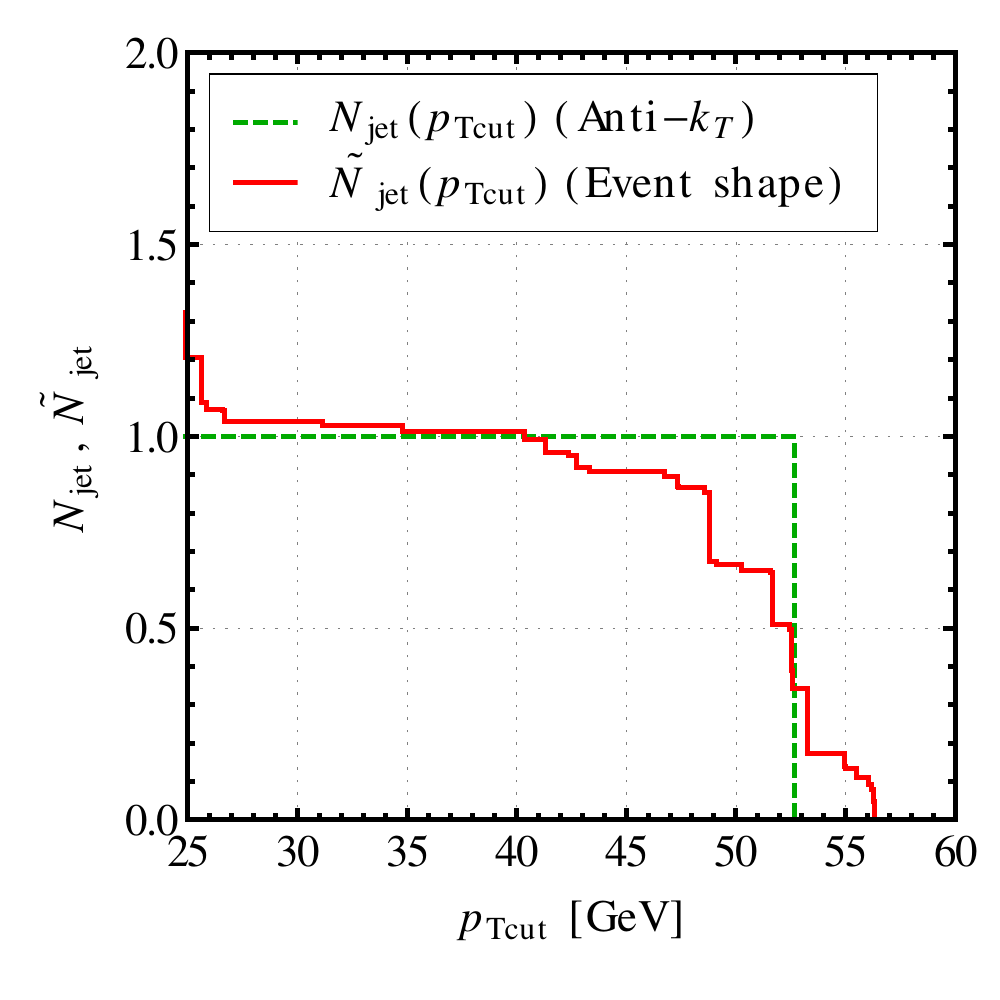}\label{fig:Npt1}}
  \subfloat[]{
    \includegraphics[scale=0.50]{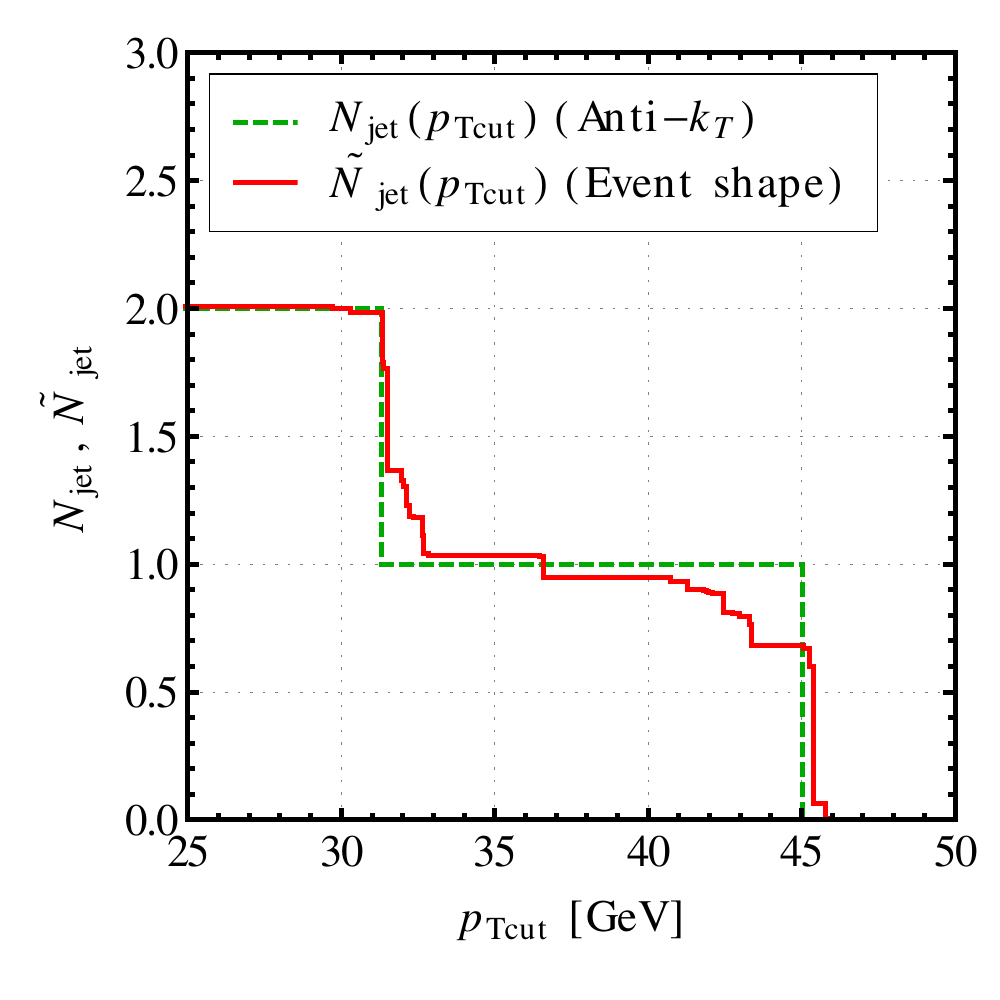}\label{fig:Npt2}}
  \subfloat[]{
    \includegraphics[scale=0.50]{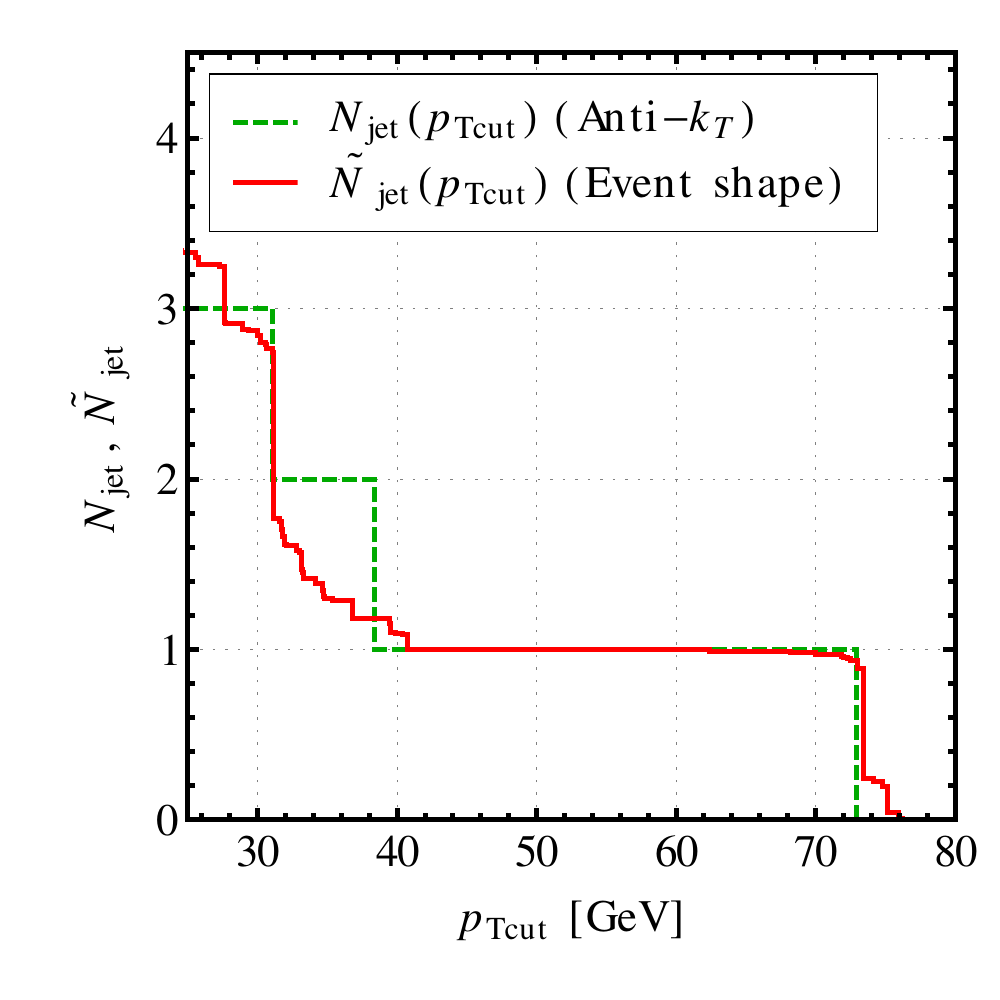}\label{fig:Npt3}}
  \caption{Number of jets $N_{\rm jet}$ as a function of $\ptc$ for fixed $R=0.6$, for three QCD dijet events. 
   \Figss{fig:Npt1}{fig:Npt2}{fig:Npt3} show example events with 1, 2, and 3 anti-$k_T$ jets with $p_T>25$~GeV, respectively.  The anti-$k_T$ curve (green dashed line) takes integer steps at values of $\ptc$ corresponding to the $p_T$ of the jets. The event shape curve (red line) takes smaller steps, and it roughly intersects the anti-$k_T$ curve at $\widetilde{N}_\jet = \{0.5,1.5,2.5\}$.}
  \label{fig:NpTplot}
 \end{figure}
 
\begin{figure}[t]
  \centering
  \subfloat[]{\includegraphics[scale=0.5]{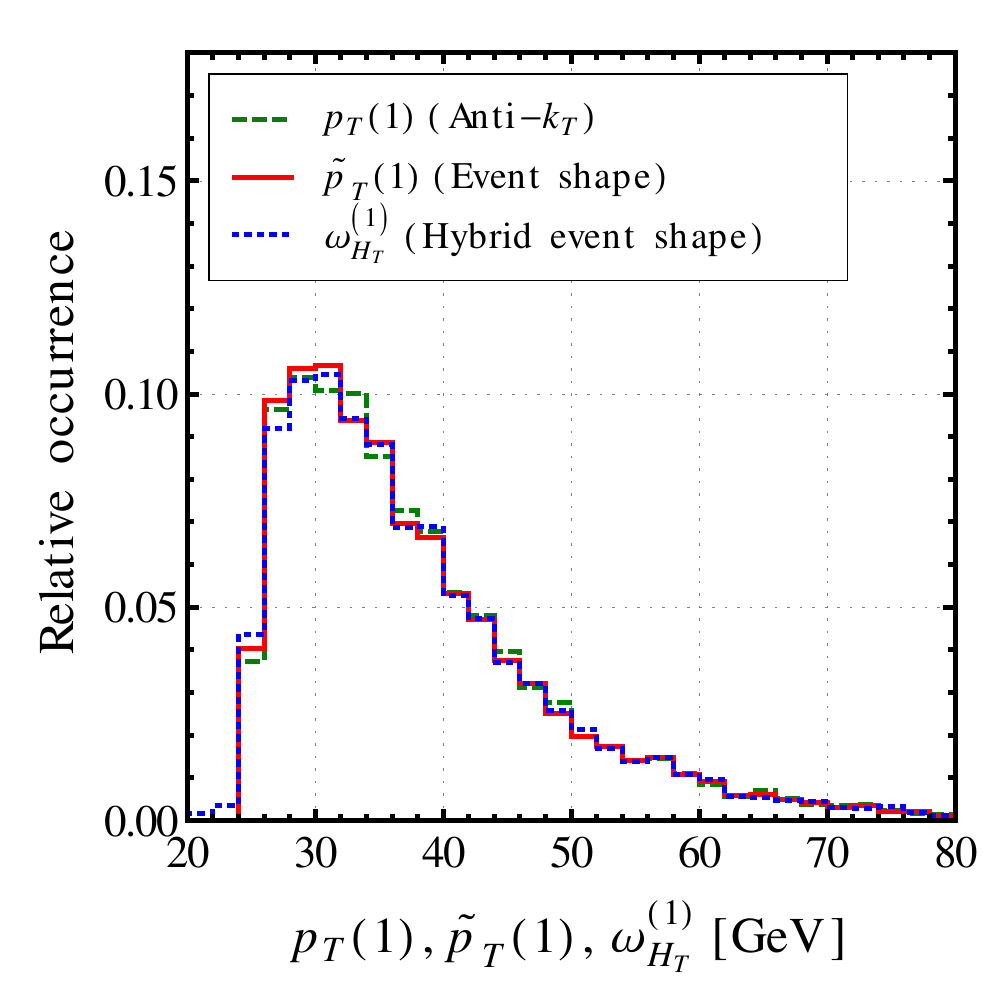}}
  \subfloat[]{\includegraphics[scale=0.5]{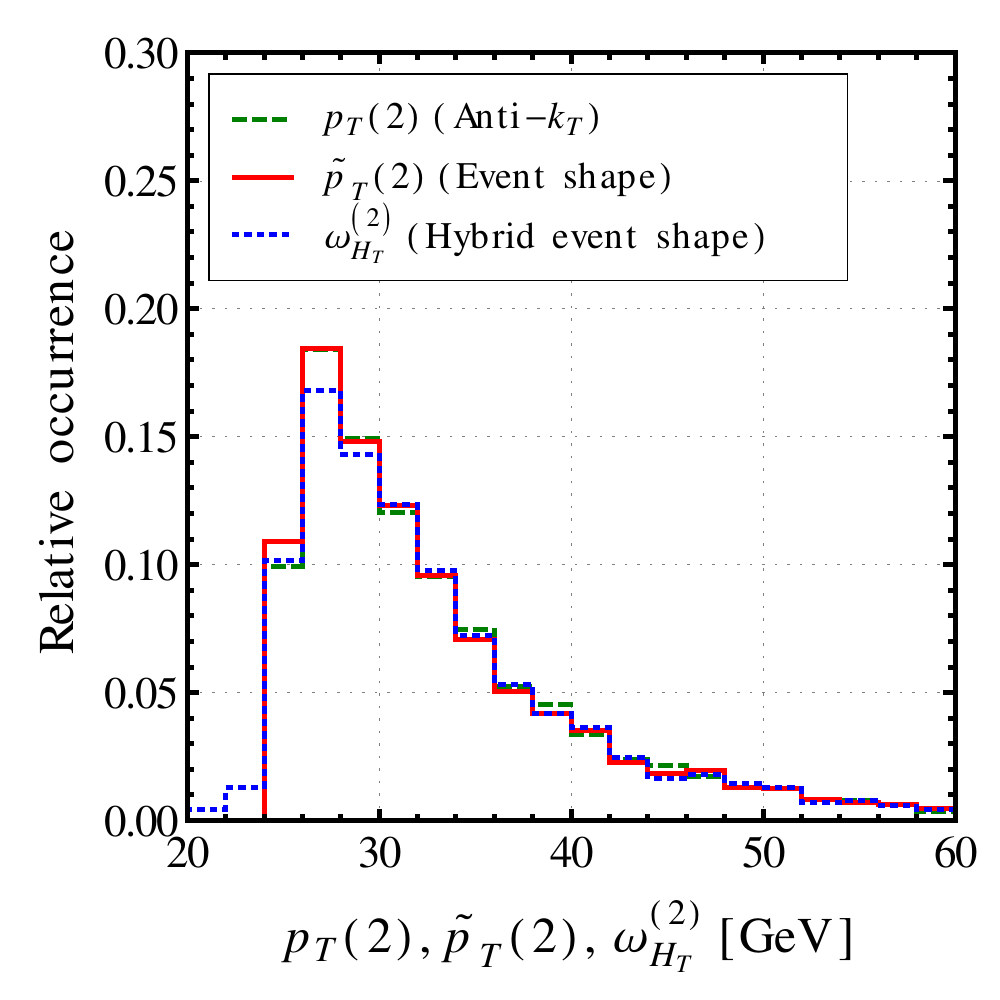}}  
  \subfloat[]{\includegraphics[scale=0.5]{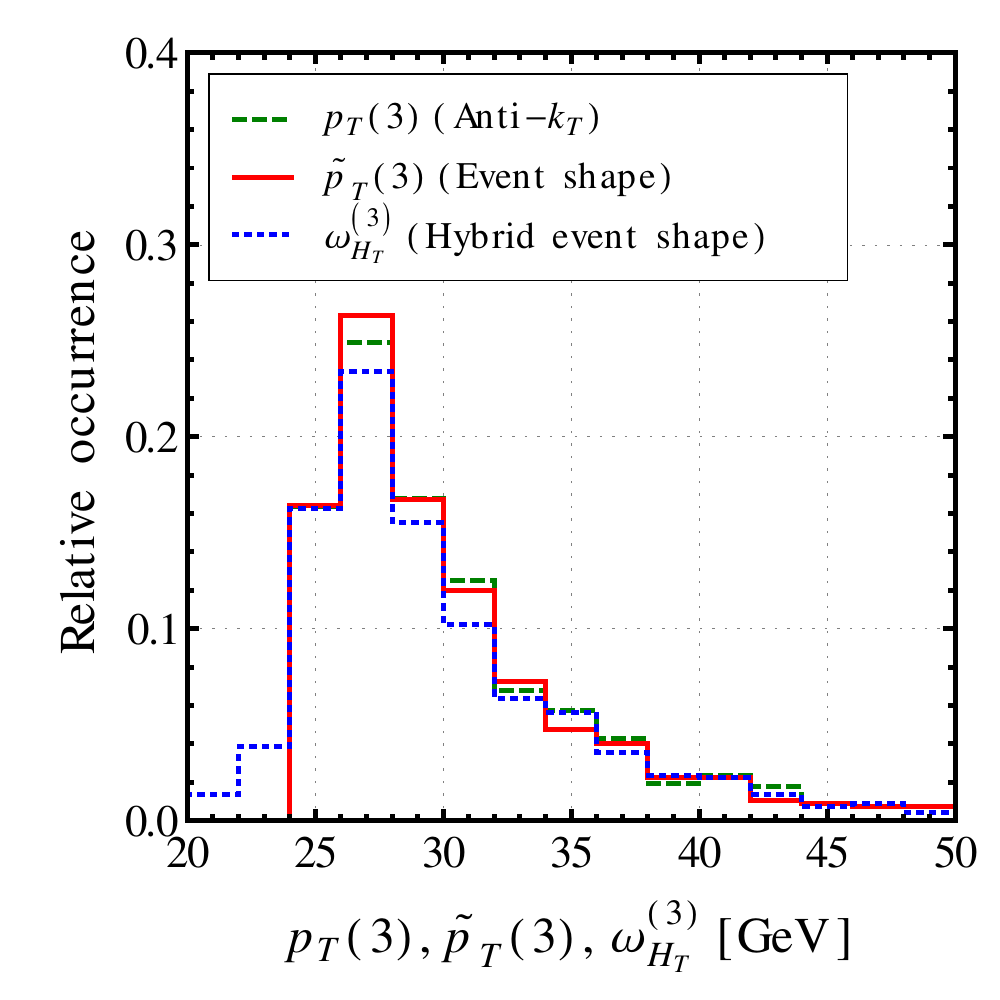}} \\
    \subfloat[]{\includegraphics[scale=0.5]{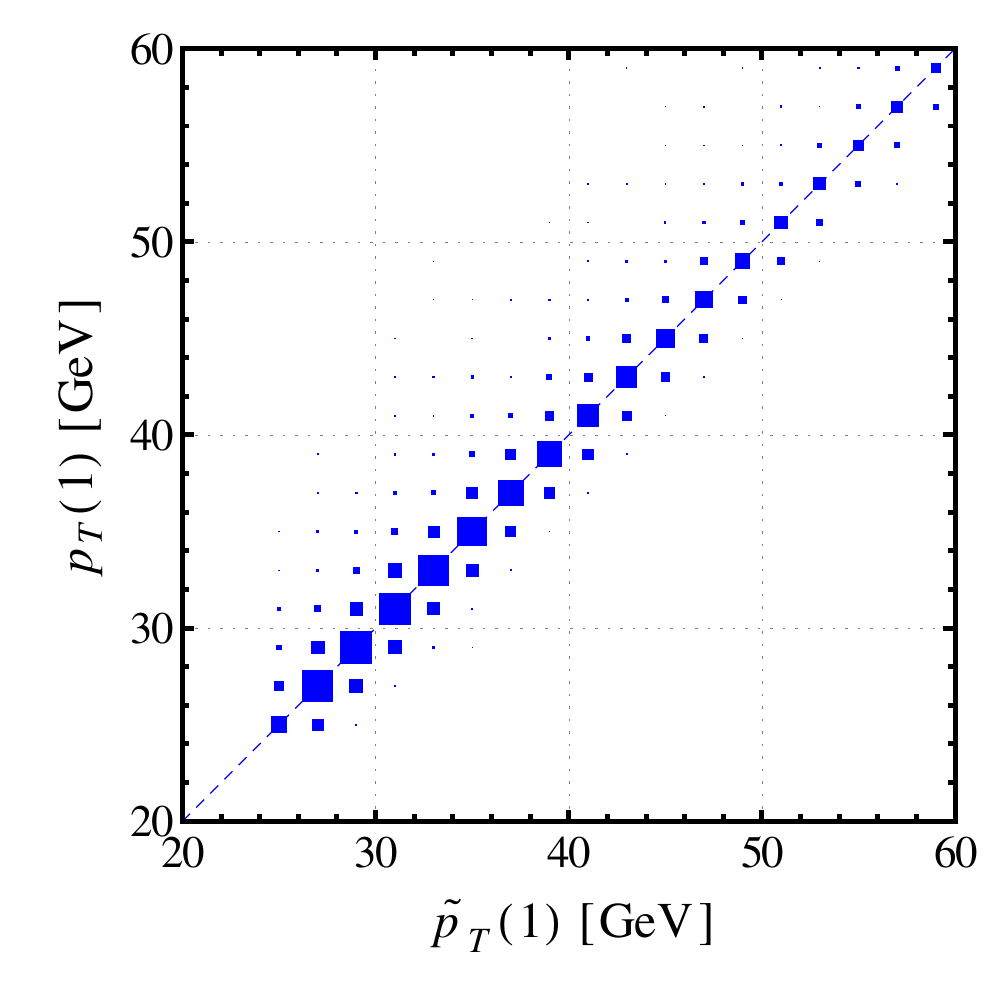}}
  \subfloat[]{\includegraphics[scale=0.5]{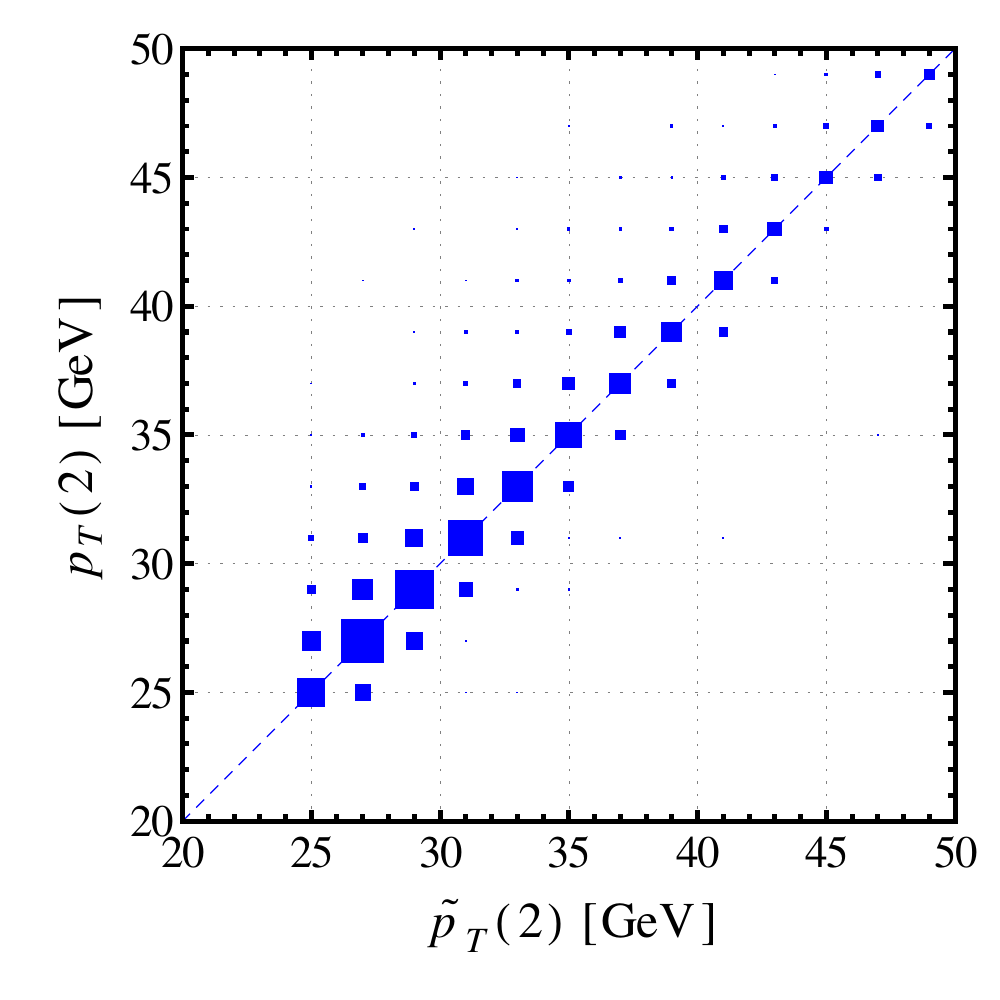}}
    \subfloat[]{\includegraphics[scale=0.5]{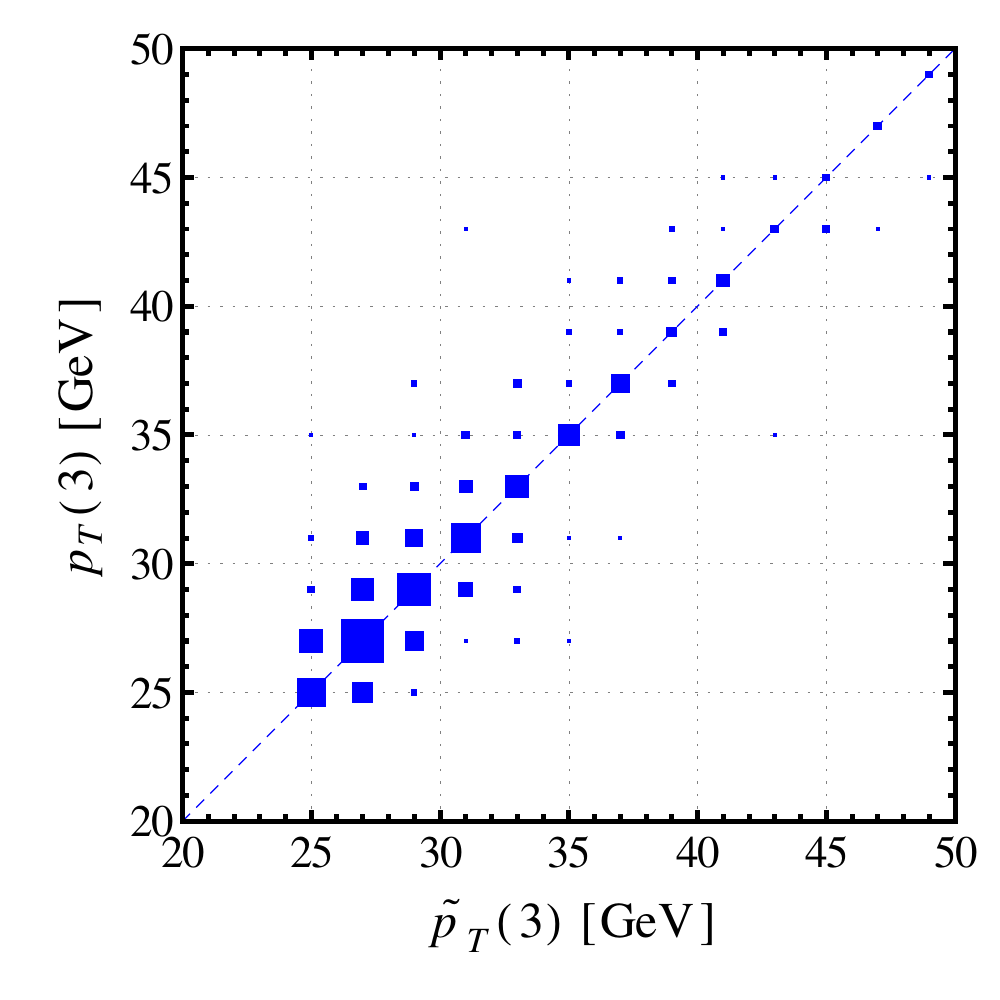}}
  \caption{Transverse momentum of the three hardest jets (i.e.\ $p_T(1)$, $p_T(2)$, and $p_T(3)$ from left to right) for QCD dijet events. The top panels shows the transverse momentum distributions for anti-$k_T$ jets with $R=0.6$ and $\ptc=25$~GeV (green dashed  curve), the corresponding event shape $\widetilde{p}_T(n)$ with the same $R$ and $\ptc$ (red curve), and the weights $\omega_{H_T}^{(n)}$ returned by the hybrid event shape with the same $R$ but $\ptc = 0$ (purple dotted curve, see \Sec{subsec:jetAxis}). The bottom panels shows the correlations between $p_T(n)$ and $\widetilde{p}_T(n)$, with the area of the squares proportional to the fraction of events in each bin.  For plots of the \{\text{1st,\,2nd,\,3rd}\}-hardest jets, the corresponding selection criteria are $N_\jet\geq 1,2,3$ (for anti-$k_T$) and $\widetilde{N}_\jet\geq 0.5,1.5,2.5$ (for the event shape).}
  \label{fig:pTnthjetplot}
 \end{figure}

Using the same QCD dijet event samples as in \Sec{sec:JetObs}, we can see how well $\widetilde{p}_T(n,R)$ corresponds to $p_T(n,R)$.  First in \Fig{fig:NpTplot}, we show the function $\widetilde{N}_\jet(\ptc,R)$ for individual events compared to $N_\jet(\ptc,R)$, fixing $R = 0.6$.  Besides the obvious point that $N_\jet$ takes integer steps whereas $\widetilde{N}_\jet$ takes smaller steps, we see that the curves roughly intersect at values of $\widetilde{N}_\jet = 0.5,1.5, 2.5$, justifying the default value $n_{\rm off} = 0.5$.  In \Fig{fig:pTnthjetplot}, we compare $\widetilde{p}_T(n,R)$ versus $p_T(n,R)$ for $n = 1,2,3$, where we see that they are highly correlated, as expected from the correlations already seen in the inclusive observables in \Sec{sec:JetObs}.

Besides just measuring the $p_T$ of the $n$-th hardest jet, $\widetilde{p}_T(n,R)$ can be used to mimic analyses that require a fixed number of jets.  For example, one may wish to measure $H_T$ on just the $n$ hardest jets above a given $\ptc$.  To do that with the event shape, one has to find the value of a new scale $\ptc'$ such that $(n+1)$-th jet would not contribute to $\widetilde{H}_T$ but the $n$-th jet is largely unmodified.  A convenient choice for that scale is
\be
\label{eq:ptcprime}
\ptc' = \max\{\ptc,\widetilde{p}_T(n+1)\},
\ee
and we will use $\ptc'$ in some of the studies in \Sec{sec:Generalizations}.

\begin{figure}
\centering
\includegraphics[scale=0.6]{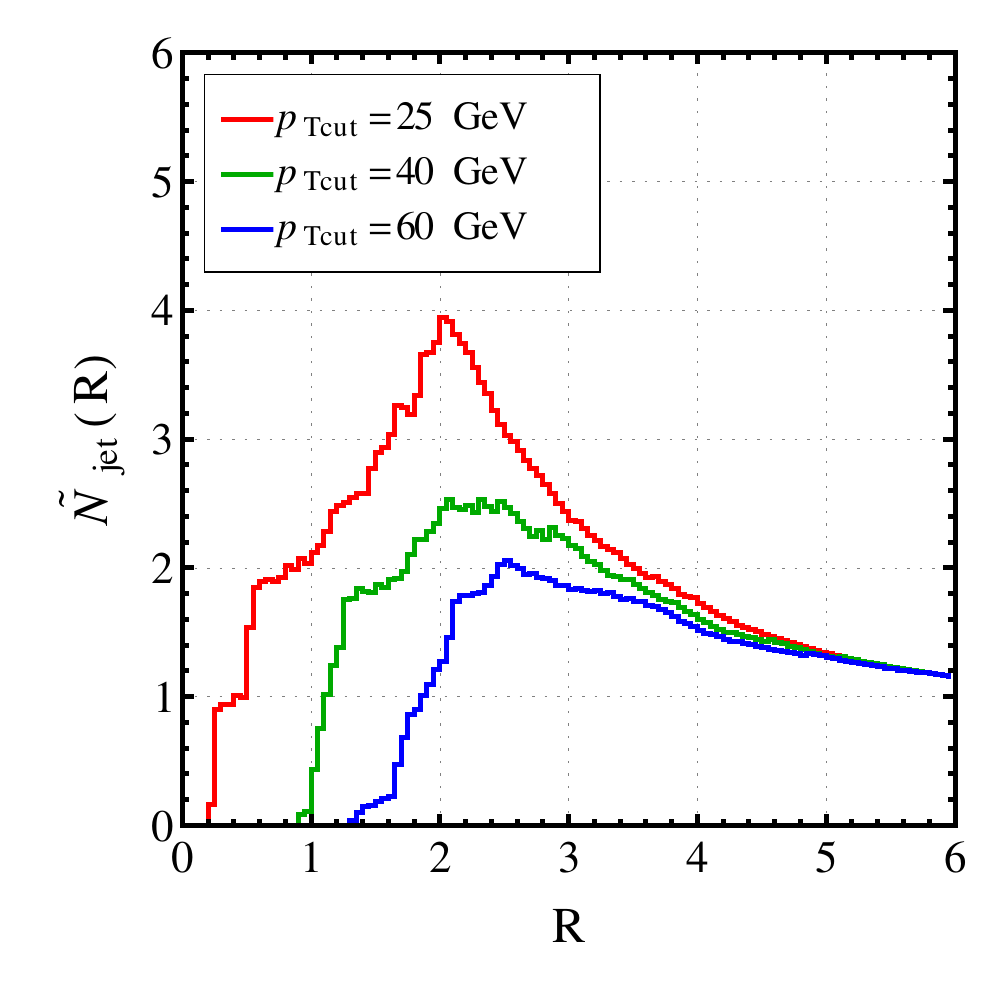}   
\caption{Number of jets $\widetilde{N}_\jet$ as a function of $R$ for a single QCD dijet event.  Shown are three values of $\ptc = \{25~\GeV, 40~\GeV, 60~\GeV\}$.}
\label{fig:NR}
\end{figure}

By using an algorithm similar to the one described in \App{app:invert}, one could also try to invert the number of jets $\widetilde{N}_\jet(R)$ as a function of $R$, for fixed $\ptc$.  Strictly speaking this inverse is not possible, since $\widetilde{N}_\jet(R)$ is not guaranteed to be a monotonic function of $R$.  Still, we expect that the $R$ dependence of the event shapes could be exploited much in the same way as for telescoping jets \cite{Chien:2013kca}.  For example, one could measure the volatility of an event shape (\`{a} la $Q$-jets \cite{Ellis:2012sn,Kahawala:2013sba}) as $R$ is varied. A detailed study of $R$ dependence is beyond the scope of this work, but in \Fig{fig:NR} we show an example of $\widetilde{N}_\jet(R)$ for a QCD dijet event, which suggests that there is interesting information to be gained by looking at multiple $R$ values.

\subsection{Jet Axes and Constituents}
\label{subsec:jetAxis}

By themselves, the event shapes do not have a clustering interpretation, so in order to (uniquely) assign particles to jets we will build a hybrid event shape that incorporates some kind of clustering procedure.  Before doing that, though, it is helpful to introduce the concept of an ``event shape density''.  

Consider the following probability density for a jet axis to lie in a given direction $\hat{n}$, as determined by a standard jet clustering algorithm:
\be
\label{eq:rhoNjet}
\rho_{N_{\rm jet}}(\hat{n})=\sum_{\jets} \delta(\hat{n}-\hat{n}^r_\jet)\,\Theta(p_{T\jet}-\ptc),
\ee
where the superscript $r$ reminds us that we must choose a recombination scheme for defining the jet axis $\hat{n}^r_\jet$ in terms of the constituents of that jet.  For example, in the standard $E$-scheme, the jet axis lies in the direction of the summed constituent four-momenta.  The reason $\rho_{N_{\rm jet}}$ is a density is that if we integrate over all directions $\hat{n}$ then $\int d^2\hat{n} \, \rho_{N_{\rm jet}} = N_{\rm jet}$, but $\rho_{N_{\rm jet}}$ itself has delta function spikes at the jet locations $\hat{n}^r_\jet$ identified by the jet algorithm.  Similarly, we can define a transverse momentum density,
\be
\rho_{H_T}(\hat{n})=\sum_{\jets} p_{T\jet} \, \delta(\hat{n}-\hat{n}_\jet)\,\Theta(p_{T\jet}-\ptc),
\ee
where $\int d^2\hat{n} \, \rho_{H_T} = H_T$ and the height of the delta functions correspond to the $p_T$ of the corresponding jets.  

Following the general strategy outlined in \Sec{sec:JetObs},  we can define corresponding event shape densities:
\begin{align}
\label{eq:rhot}
\widetilde{\rho}_{N_{\rm jet}}(\hat{n}) &=\sum_{i\in\ev}\frac{p_{Ti}}{p_{Ti,R}}\delta(\hat{n}-\hat{n}_{i,R}^r)\,\Theta(p_{Ti,R}-\ptc), \\
\label{eq:rhoHT}
\widetilde{\rho}_{H_T}(\hat{n}) &=\sum_{i\in\ev} p_{Ti}\, \delta(\hat{n}-\hat{n}_{i,R}^r)\,\Theta(p_{Ti,R}-\ptc),
\end{align}
where $\int d^2\hat{n} \, \widetilde{\rho}_{N_{\rm jet}} = \widetilde{N}_{\rm jet}$ and $\int  d^2\hat{n} \, \widetilde{\rho}_{H_T} = \widetilde{H}_T$.  Here, $\hat{n}_{i,R}^r$ is the direction of the recombined momenta in a cone of radius $R$ around particle $i$, which of course depends on the recombination scheme $r$.  If we choose to do recombination via the $E$-scheme, then $\widetilde{\rho}_{N_{\rm jet}}$ and $\widetilde{\rho}_{H_T}$ can still be considered event shapes, since $\hat{n}_{i,R}^r$ can be written in closed form (i.e.\ in terms of the four-vector sum of constituents).  For more general recombination schemes, though, $\widetilde{\rho}_{N_{\rm jet}}$ and $\widetilde{\rho}_{H_T}$ are hybrid event shapes, since the specific direction of $\hat{n}_{i,R}^r$ depends on the recombination algorithm (which in general cannot be written in closed form).  In contrast to standard jet clustering algorithms, finding $\hat{n}_{i,R}^r$ is a ``local'' procedure since it only requires knowledge about particles within a radius $R$ of particle $i$.

Whereas the jet-based densities have $n$ delta function spikes for an $n$-jet event, the event shape densities typically exhibit a more continuous distribution. In particular, the distribution will still show peaks corresponding to jet directions, although smeared because nearby particles will typically have (slightly) different values of $\hat{n}_{i,R}^r$.  In this way, the event shape densities are similar in spirit to the jet energy flow project \cite{Berger:2002jt}, since they effectively give a probability distribution for the jet axis locations.  

Concretely, if we let $\{\hat{m}_j^r\}$ be the set of distinct directions in $\{\hat{n}_{i,R}^r\}$, we can rewrite the distributions in \Eqs{eq:rhot}{eq:rhoHT} as
\be
\label{eq:rhot_prime}
\widetilde{\rho}_X(\hat{n})=\sum_{j} \omega_{Xj} \,\delta(\hat{n}-\hat{m}_j^r),\quad X=N_\jet,H_T.
\ee
The coefficients $\omega_{Xj}$ can be thought as weights corresponding to each candidate jet axis $\hat{m}_j^r$ and are given by: 

\be
\label{eq:weights}
\begin{split}
\omega_{N_\jet j}&=\sum_{i\in\ev}\frac{p_{Ti}}{p_{Ti,R}}\,\Theta(p_{Ti,R}-\ptc)\,\delta_{\{\hat{n}_{i,R}^r;\,m_j^{r}\}},\\
\omega_{H_Tj}&=\sum_{i\in\ev}p_{Ti}\,\Theta(p_{Ti,R}-\ptc)\,\delta_{\{\hat{n}_{i,R}^r;\,m_j^{r}\}},
\end{split}
\ee
where $\delta_{\{\hat{n}_{i,R}^r;\,m_j^{r}\}}$ is a Kronecker delta over the discrete sets of directions $\{\hat{n}_{i,R}^r\}$ and $\{m_j^{r}\}$. The weights $\omega_{N_{\rm jet}j}$ indicate the (fractional) number of jets that should be associated with a given axis, while $\omega_{H_Tj}$ indicate the associated transverse momentum.  For an isolated narrow jet, a typical recombination scheme will yield a single axis $\hat{m}^r$ with $\omega_{N_{\rm jet}} = 1$ and $\omega_{H_T} = p_{T\jet}$.

We emphasize that in this hybrid approach, a separate clustering algorithm is applied to each particle $i$, using just the particles within its neighborhood of radius $R$.  For an event with $N$ final state hadrons, one has to run $N$ clustering algorithms, yielding $N$ values of $\hat{n}_{i,R}^r$, though not all of them will be distinct.  In practice, it is inconvenient to have $\mathcal{O}(N)$ candidate jet axis locations, so ideally we want a recombination scheme that returns $\mathcal{O}(n)$ unique axes $\hat{m}_j^r$ for an $n$-jet event. 

For this purpose, we will use a ``winner-take-all'' recombination scheme when performing the local clustering around each particle.\footnote{We thank Andrew Larkoski, Duff Neill, and Gavin Salam for discussions on this point.  The winner-take-all scheme is also discussed in \Ref{Larkoski:2014uqa} in the context of recoil-free observables.}  This scheme guarantees that the recombined direction will always coincides with one of the input particles, dramatically decreasing the number of unique $\hat{m}_j^r$ values.  In the context of a pairwise clustering algorithm like anti-$k_T$, the recombination scheme determines how two pseudo-jets $p_1$ and $p_2$ will be merged to form a combined pseudo-jet $p_r$.  In the winner-take-all scheme, the transverse momentum of $p_r$ is given by the sum of the two pseudo-jets, but the direction of $p_r$ is given by the hardest pseudo-jet:
\be
p_{Tr} = p_{T1} + p_{T2}, \qquad \hat{n}_r =  \begin{cases} \hat{n}_1 & \text{if $p_{T1} > p_{T2}$}, \\
\hat{n}_2 & \text{if $p_{T2} > p_{T1}$}. \end{cases}
\ee
For simplicity, we take $p_r$ to be a massless four-vector.  When used with an infrared/collinear safe clustering measure (anti-$k_T$ in the later plots), the winner-take-all scheme is also infrared/collinear safe.  Because the winner-take-all scheme always returns a jet direction aligned along one of the input particles (often the hardest particle), the set of recombined jet directions $\{\hat{m}_j^r\}$ is much smaller than the number of hadrons in the final state.\footnote{To further reduce the number of jet directions, we could further insist that the winner-take-all axes are globally consistent.  That is, if particle $a$ has winner-take-all axis aligned with particle $b$, but particle $b$ has winner-take-all axis aligned with particle $c$, then we could assign particle $a$ the axis aligned with $c$ (recursing further if necessary).  This consistency criteria would ensure that the final set of jet directions $\{\hat{m}_j^r\}$ are their own winner-take-all axes.  It would also imply that the jet regions can expand beyond a cone of radius $R$ from the jet axes.  This option is available in the \FastJet\ add-on, but not used in the following plots.}  Of course, for later analysis, one probably wants to use the summed four-vector of the jet constituents instead of the jet axis.\footnote{Unlike in the $E$-scheme, the jet axis and the jet four-momentum (i.e.~the summed four-momenta of the jet constituents) will not typically be aligned in the winner-take-all scheme.} 

\begin{figure}[t]
  \centering
  \subfloat[]{
    \includegraphics[scale=0.71]{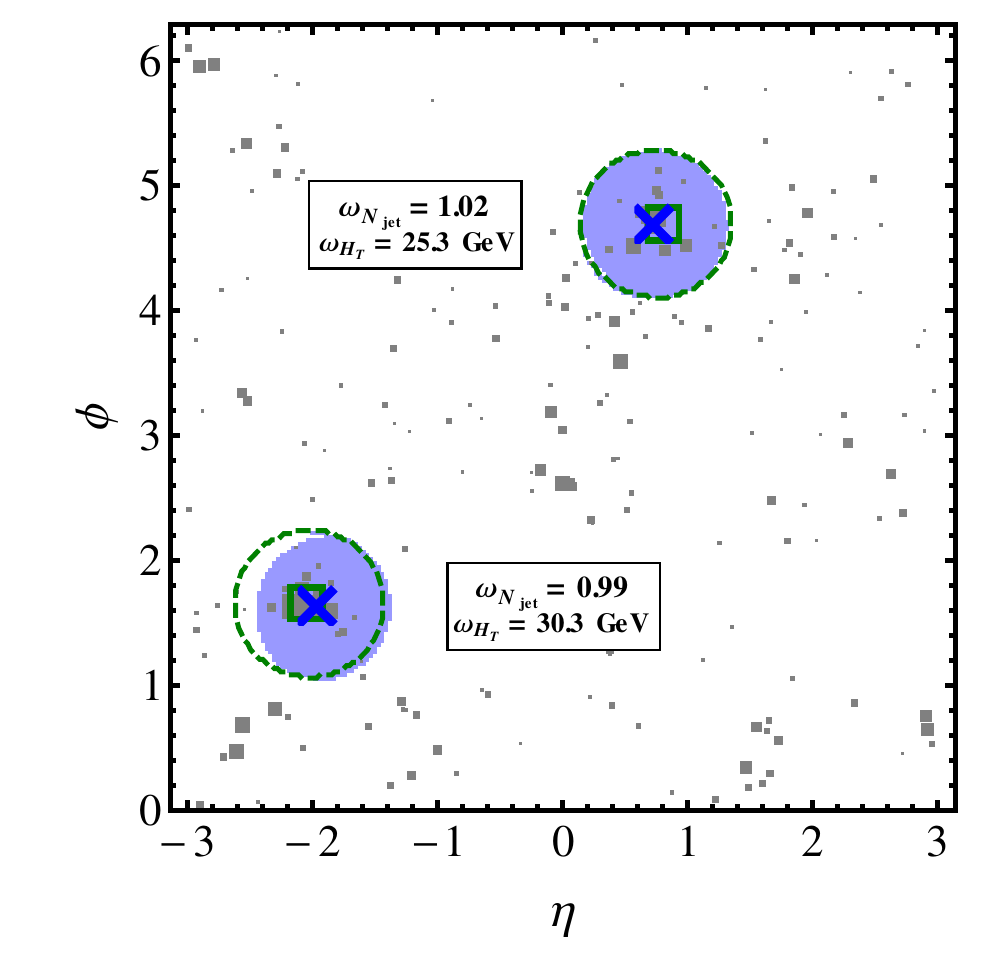}
}   
\hspace{0.1in}
  \subfloat[]{
      \includegraphics[scale=0.71]{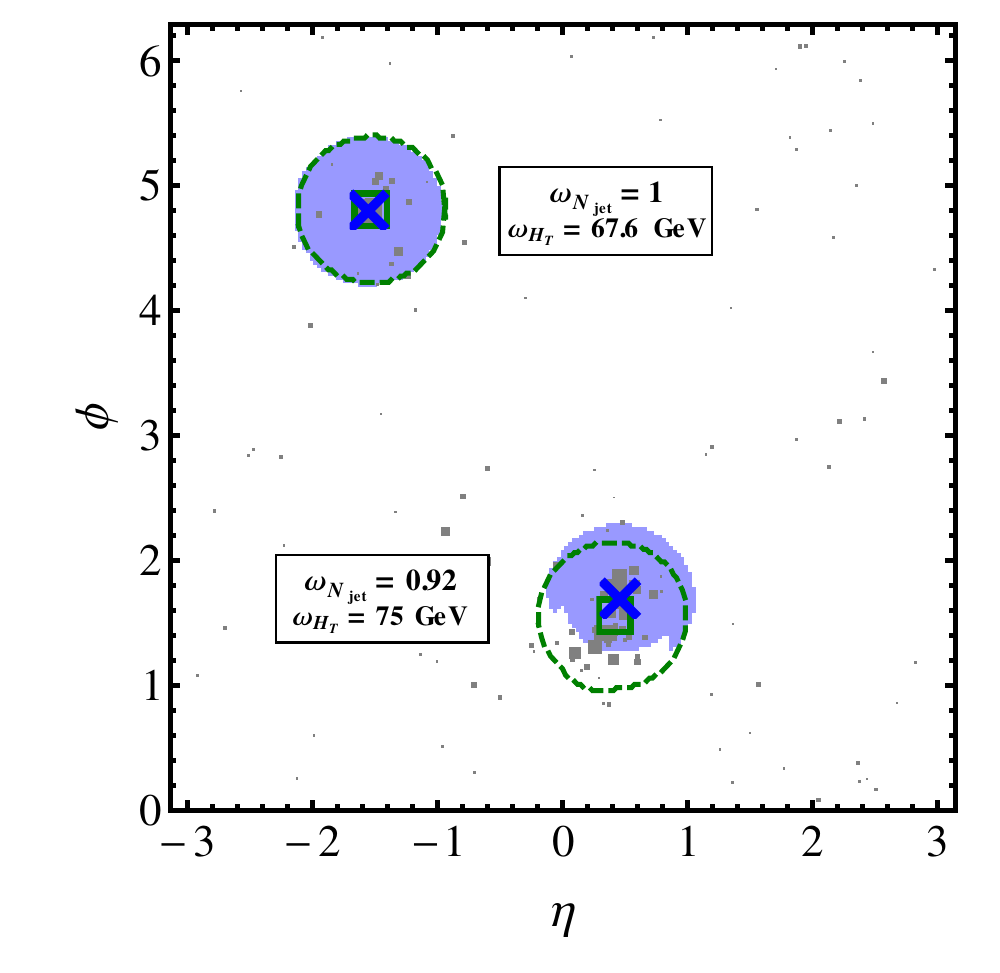}
}
 \caption{Two QCD dijet event displays with $R=0.6$. The anti-$k_T$ jet axes (green) are compared to the ones obtained using the hybrid event shape approach with $\ptc = 0$ (blue).   The standard $E$-scheme is used for the anti-$k_T$ jets, whereas the hybrid event shape uses the winner-take-all recombination scheme, as explained in \Sec{subsec:jetAxis}.  The light blue shaded region corresponds to (passive) ghost particles which are clustered to the given axis, and the dashed green curve gives the anti-$k_T$ boundary. The weights $\omega_{N_{\rm jet}} \simeq 1$ and $\omega_{H_T} \simeq p_{T\jet}$ associated with the event shape axes are also shown.}
  \label{fig:EventDisplay}
 \end{figure}

Another practical consideration concerns the value of $\ptc$. As stated above, one can think of $\omega_{H_Tj}$ in \Eq{eq:weights} as the transverse momentum associated with jet $j$, so that a way to find the $n$ hardest jets is by taking the $n$ highest values of $\omega_{H_T}$. However, although the sum of the $\omega_{H_T}$ returns $\widetilde{H}_T$, $\ptc$ would distort the jet $p_T$ spectrum.  The reason is that the $\ptc$ requirement in \Eqs{eq:rhot}{eq:rhoHT} vetoes particles near the periphery of jets which would be captured using standard clustering procedures.  Note that this effect is relevant only for jets close to the $\ptc$ threshold.  This effect was not seen in \Fig{fig:pTnthjetplot} for $\widetilde{p}_T(n)$ because there we could compensate for the loss of peripheral particles by using $n_{\text{off}} = 0.5$ in \Eq{eq:noffset}.  This effect is visible, however, in \Fig{fig:HistoHT} for $H_T$ where the peaks in the event shape $\widetilde{H}_T$ (corresponding to jets at threshold) are below the peaks for the jet-based $H_T$ because of leakage towards smaller values of $\widetilde{H}_T$.  The most convenient way to restore the vetoed particles is to simply take $\ptc= 0$ in \Eq{eq:weights}, in which case the sum of the $\omega_{H_T}$ yields the total sum of scalar $p_T$ in the event (though the sum of the $\omega_{N_\jet}$ is no longer infrared safe).

We now compare standard jet clustering to the hybrid event shape approach. For anti-$k_T$ jets, we use the standard $E$-scheme recombination, whereas for the hybrid event shape, we use the anti-$k_T$ clustering measure with winner-take-all recombination for the local clustering around each particle.  In \Fig{fig:EventDisplay} we show two QCD dijet events comparing the two hardest jets from anti-$k_T$ with the jets defined by the two highest weights $\omega_{H_T}$ (with $\ptc = 0$).  We also show the corresponding values of $\omega_{N_\jet}$ and $\omega_{H_T}$.  The displayed jet regions are determined by adding (passive) ghost particles \cite{Cacciari:2008gn}. There are differences between the jet axes caused by the different recombination schemes, and differences in the jet regions from the different effective jet splitting criteria.  But overall, there is a good correlation between the two methods, and the fact that $\omega_{N_\jet} \simeq 1$ is a nice cross check.

\begin{figure}[p]
  \centering
  \subfloat[]{\label{fig:ScatterEta}
    \includegraphics[scale=0.5]{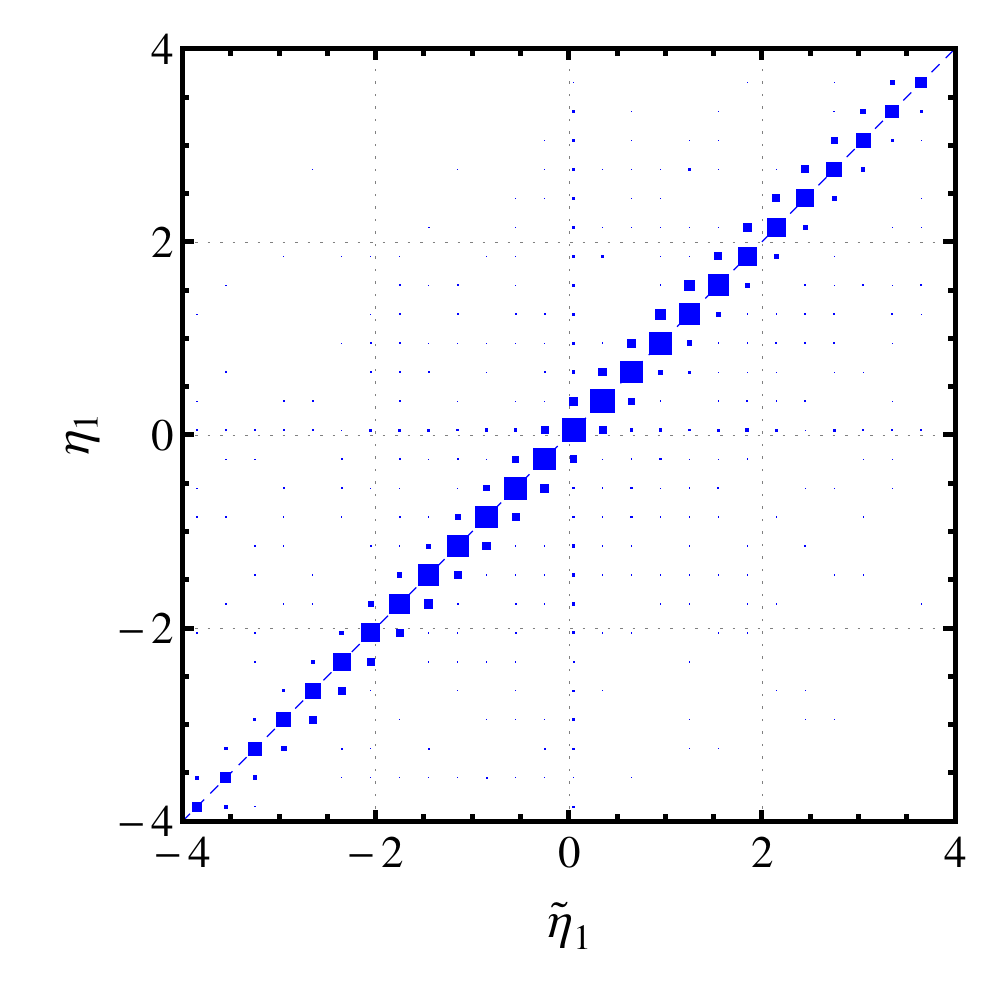}}   
\hspace{0.3in}
  \subfloat[]{\label{fig:ScatterPhi}
    \includegraphics[scale=0.5]{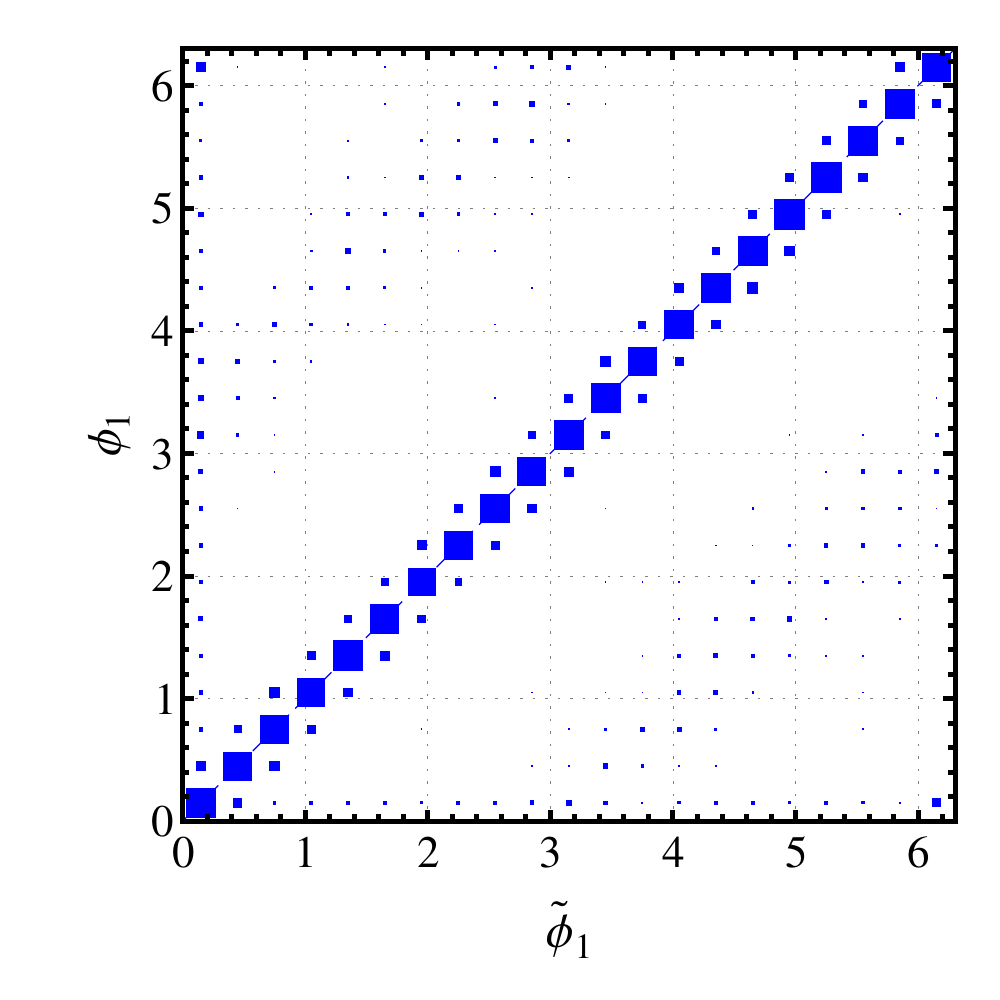}}
  \caption{Position of the hardest jet axis for QCD dijet events, using the same jet clustering as \Fig{fig:EventDisplay}.  The $(\eta_1,\phi_1)$ coordinates correspond to the jet axis identified with anti-$k_T$, and $(\widetilde{\eta}_1,\widetilde{\phi}_1)$ are the coordinates found with the hybrid event shape.  The area of the squares is proportional to the fraction of events in each bin.  There is a slight difference in the jet direction due to the different recombination schemes ($E$-scheme for anti-$k_T$, winner-take-all for the hybrid event shape).  Note the (small) accumulation of events at $|\phi_1 - \widetilde{\phi}_1| = \pi$, which occur when the two algorithms disagree about which of the dijets is the hardest.}
\label{fig:AxesCorrelation}
  \end{figure}

\begin{figure}[p]
\centering
 \subfloat[]{\label{fig:HistoJetAreas}
   \includegraphics[scale=0.5]{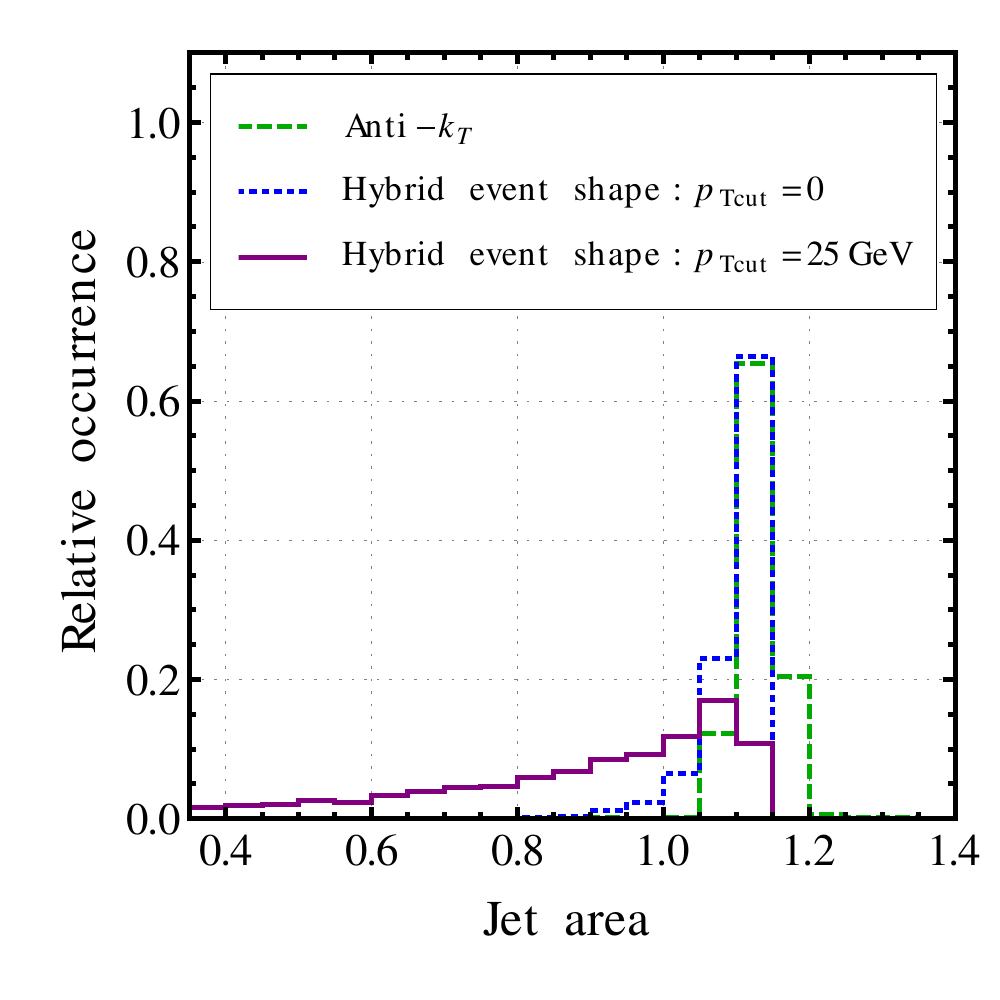}}   
  \subfloat[]{\label{fig:HistoWHT}
    \includegraphics[scale=0.5]{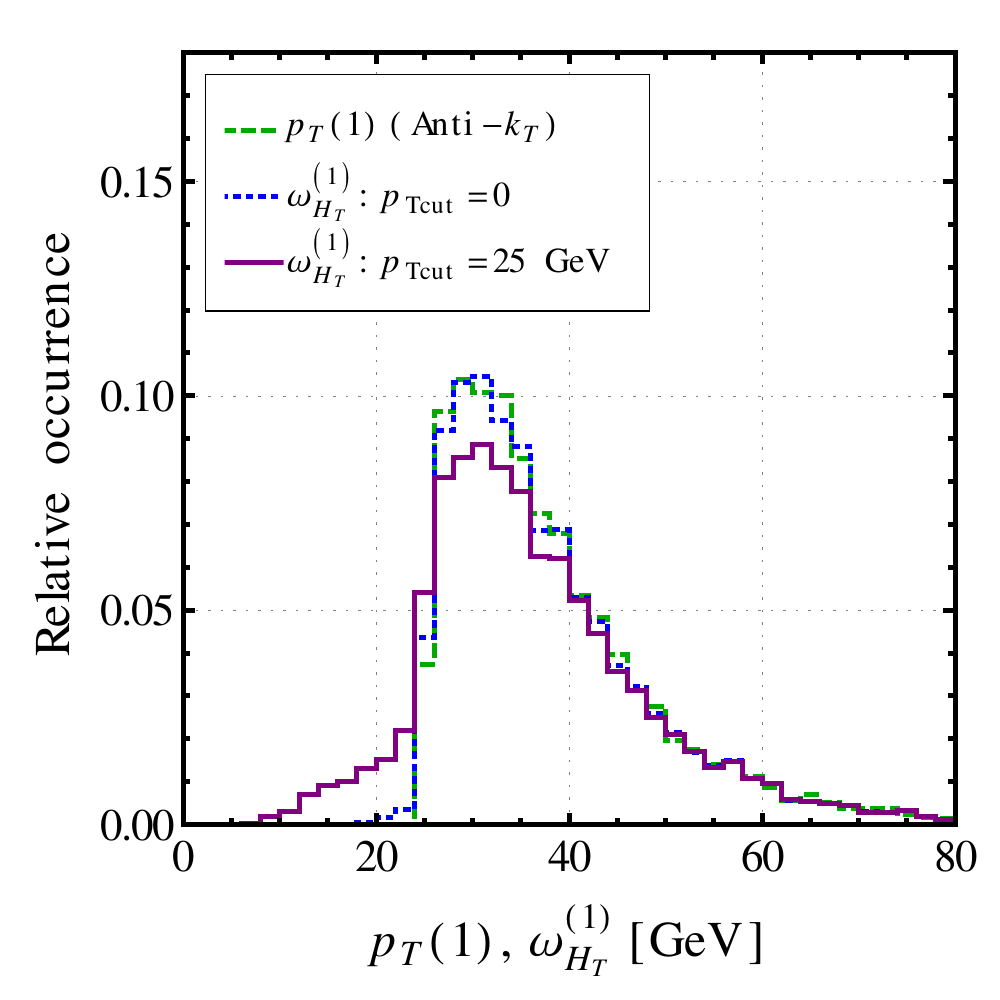}
    }
  \subfloat[]{\label{fig:HistoWNjet}
    \includegraphics[scale=0.5]{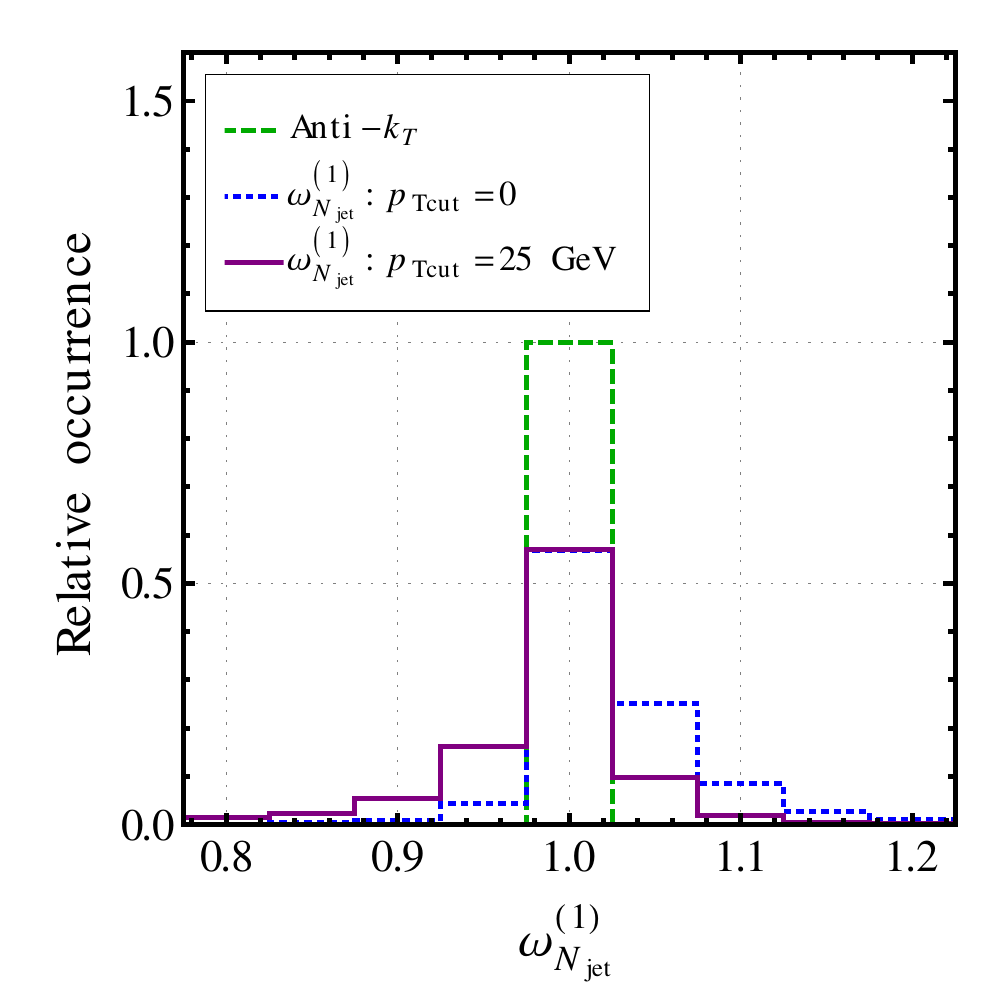}}
  \caption{Comparison of the hardest jet found with anti-$k_T$ (green dashed curve), the hybrid event shape result with $\ptc=0$ (blue dotted curve), and the hybrid event shape with $\ptc=25$~GeV (purple curve), all for QCD dijet events with $R=0.6$.  Left:  Passive jet area, where the first two methods peak at $\pi R^2$.  Center:  jet $p_T$ (or $\omega_{H_T}$).  Right:  fractional jet weight $\omega_{N_\jet}$, where all methods peak at 1.  In all cases, the $\ptc=0$ event shape is closer to the anti-$k_T$ result, since it restores peripheral particles that are vetoed with non-zero $\ptc$.}
   \label{fig:jetAreasAndWeights}
  \end{figure}

Turning to the QCD dijet event sample, back in \Fig{fig:pTnthjetplot} we showed distributions for the three highest weights $\omega_{H_T}$, which correlate strongly with the three hardest jets from anti-$k_T$ (and with the inverse multiplicity $\widetilde{p}_T(n)$).  In \Fig{fig:AxesCorrelation}, we compare the direction of the axis of the hardest jet found with both methods, again seeing good agreement, apart from a small set of events where the azimuth differs by $\pi$ because the choice of hardest jet is ambiguous.  In the three panels of \Fig{fig:jetAreasAndWeights}, we show various effects on the hardest jet of having $\ptc = 0$ versus non-zero $\ptc$.   The (passive) jet areas are shown in \Fig{fig:HistoJetAreas}, where  the jet area distribution is peaked around $\pi R^2$ for $\ptc = 0$ (similar to anti-$k_T$) whereas the area is smaller for non-zero $\ptc$ because of peripheral vetoes.  The same effect is seen in \Fig{fig:HistoWHT}, where a non-zero $\ptc$ decreases the $\omega_{H_T}$ value.  The effect is less visible for $\omega_{N_\jet}$ in the \Fig{fig:HistoWNjet}, since most events peak at 1, but there is a shift to lower $\omega_{N_\jet}$ as $\ptc$ is increased.  We thus conclude that $\ptc = 0$ gives results that are closer to the expectation from standard jet clustering.

In terms of computational cost, the hybrid event shape approach is significantly more costly than anti-$k_T$, since one has to effectively run a separate jet clustering procedure for each particle $i$ to determine the direction $\hat{n}_{i,R}^r$.  On a standard laptop, it is roughly a factor of four slower on dijet events.  Despite the speed issue, this approach to identifying candidate jet regions might still be appropriate for trigger-level analyses because of the parallelizable and local nature of the hybrid event shapes.  The winner-take-all recombination is crucial for this approach to work, since it ensures that only a small number of candidate jet axes are identified.  It also has the nice feature that a given jet axis is guaranteed to align along one of the input particle directions.

\section{Shape Trimming}
\label{sec:Trimming}

Thus far, we have only discussed event shapes for observables built as a sum over all jets in an event.  As discussed further in \Sec{sec:Generalizations}, the same basic strategy can be applied to observables which are a sum over all subjets in all jets in an event.  A simple application of this is to implement jet trimming \cite{Krohn:2009th} via an event shape.  We refer to traditional jet trimming as ``tree trimming'' and the corresponding event shape version as ``shape trimming''.

In tree trimming, one first clusters particles into jets of radius $R$ and $p_{T\jet} > \ptc$, typically via the anti-$k_T$ algorithm.  For each jet, one reclusters its constituents into subjets with characteristic radius $R_\text{sub}<R$, typically via the CA algorithm \cite{Dokshitzer:1997in,Wobisch:1998wt,Wobisch:2000dk} or $k_T$ algorithm \cite{Catani:1993hr,Ellis:1993tq}.   Subjets whose transverse momentum fraction $p_{T\sub}/p_{T\jet}$ are above a certain threshold $f_{\text{cut}}$ are kept, while the remaining subjets are removed.  The four-momentum of a trimmed jet can be written as
\be
\label{eq:TrimJet}
t^\mu_\jet=\sum_\subjs p^\mu_\sub \Theta\left(\frac{p_{T\sub}}{p_{T\jet}}-f_\text{cut}\right),
\ee
where $p^\mu_\sub$ is the four-momentum of the subjet, $p_{T\sub}$ is the corresponding transverse momentum, and $p_{T\jet}$ is the transverse momentum of the un-trimmed jet.  The trimmed four-momentum of the entire event is
\be
\label{eq:tev}
t^\mu_\ev=\sum_\jets t^\mu_\jet \Theta(p_{T\jet}-\ptc) = \sum_\jets \sum_\subjs p^\mu_\sub \Theta\left(\frac{p_{T\sub}}{p_{T\jet}}-f_\text{cut}\right) \Theta(p_{T\jet}-\ptc).
\ee
Along with the clustering algorithms used, the trimming procedure is specified by the jet parameters $\{\ptc,R\}$ and the subjet parameters $\{f_{\text{cut}},R_\text{sub}\}$.  

To recast trimming as an event shape, we can follow the strategy outlined in \Sec{sec:JetObs}, but adding an extra step to deal with the presence of subjets.  Since $p^\mu_\sub\simeq\sum_{i\in\text{subjet}}p^\mu_i$ can be written as a sum over subjet's constituents, we can skip the first replacement in \Eq{eq:repl1}, and directly make the replacement 
\be
\sum_\jets \sum_\subjs  p^\mu_\sub \to \sum_{i \in \ev}  p^\mu_i.
\ee
Moreover, 
\be
p_{T\jet} \to p_{Ti, R}, \qquad p_{T\sub} \to p_{Ti, R_{\rm sub}},
\ee
where  $p_{T i,R_\sub}$ is analogous to $p_{T i,R}$ in \Eq{eq:pTR}, except it only includes particles contained in a cone around particle $i$ of radius $R_\sub$.  The trimmed event shape corresponding to the overall four-momentum is therefore 
\be
\label{eq:ttev}
\widetilde{t}^{\,\mu}_\ev=\sum_{i\in\ev}p^\mu_i \, \Theta\left(\frac{p_{T i,R_\sub}}{p_{T i,R}}-f_\text{cut}\right)\Theta(p_{T i,R}-\ptc).
\ee

For defining more general event shapes (or for use in other jet-based analyses), we can interpret $\widetilde{t}^{\,\mu}_\ev$ as defining a weight for each individual particle:
\be
\label{eq:trimweights}
w_i = \Theta\left(\frac{p_{T i,R_\sub}}{p_{T i,R}}-f_\text{cut}\right)\Theta(p_{T i,R}-\ptc).
\ee
Here $w_i$ is either 0 or 1, but one could generalize $w_i$ to take on continuous values by smoothing out the theta functions.  In practice, we implement \Eq{eq:trimweights} as a \texttt{Selector} in our \FastJet\ add-on, which takes a collection of particles and only returns those particles with $w_i = 1$.  Instead of applying trimming event wide (``event shape trimming''), one could first find jets with an ordinary jet algorithm and then apply \Eq{eq:trimweights} with $p_{T i,R}$ replaced by $p_{T\jet}$; we have implemented this ``jet shape trimming'' option as a \texttt{Transformer} in  \FastJet.

One could also use the weights directly in the event shapes.  For example, we could define the trimmed jet multiplicity as
\be
\label{eq:Ntrim}
\widetilde{N}^\text{trim}_\jet(\ptc,R;f_\text{cut}, R_{\sub}) = \sum_{i\in\ev}\frac{p_{T i}}{p_{T i,R}}\Theta\left(\frac{p_{T i,R_\sub}}{p_{T i,R}}-f_\text{cut}\right) \Theta(p_{T i,R}-\ptc),
\ee
and one could define the trimmed inverse $\widetilde{p}^{\,\text{trim}}_T(n,R;f_\text{cut}, R_{\sub})$ accordingly.  Note that applying the weights in \Eq{eq:trimweights} first and then calculating $\widetilde{N}_\jet$ is not the same as calculating $\widetilde{N}^\text{trim}_\jet$ directly, since in the former case, the value of $p_{T i,R}$ is affected by the weights. In most cases, one gets better performance by using the weights first, especially if the jet observable $\F_{\jet}$ is non-linear in the inputs (as is the case for jet mass studied in \Sec{subsec:singlejet}).  For $\widetilde{N}^\text{trim}_\jet$ there is only a mild difference, so we use $\widetilde{N}^\text{trim}_\jet$ for simplicity in some of the case studies in \Sec{sec:Generalizations}.

\begin{figure}[t]
  \centering
  \subfloat[]{\label{fig:TrimEvent1}
    \includegraphics[scale=0.55]{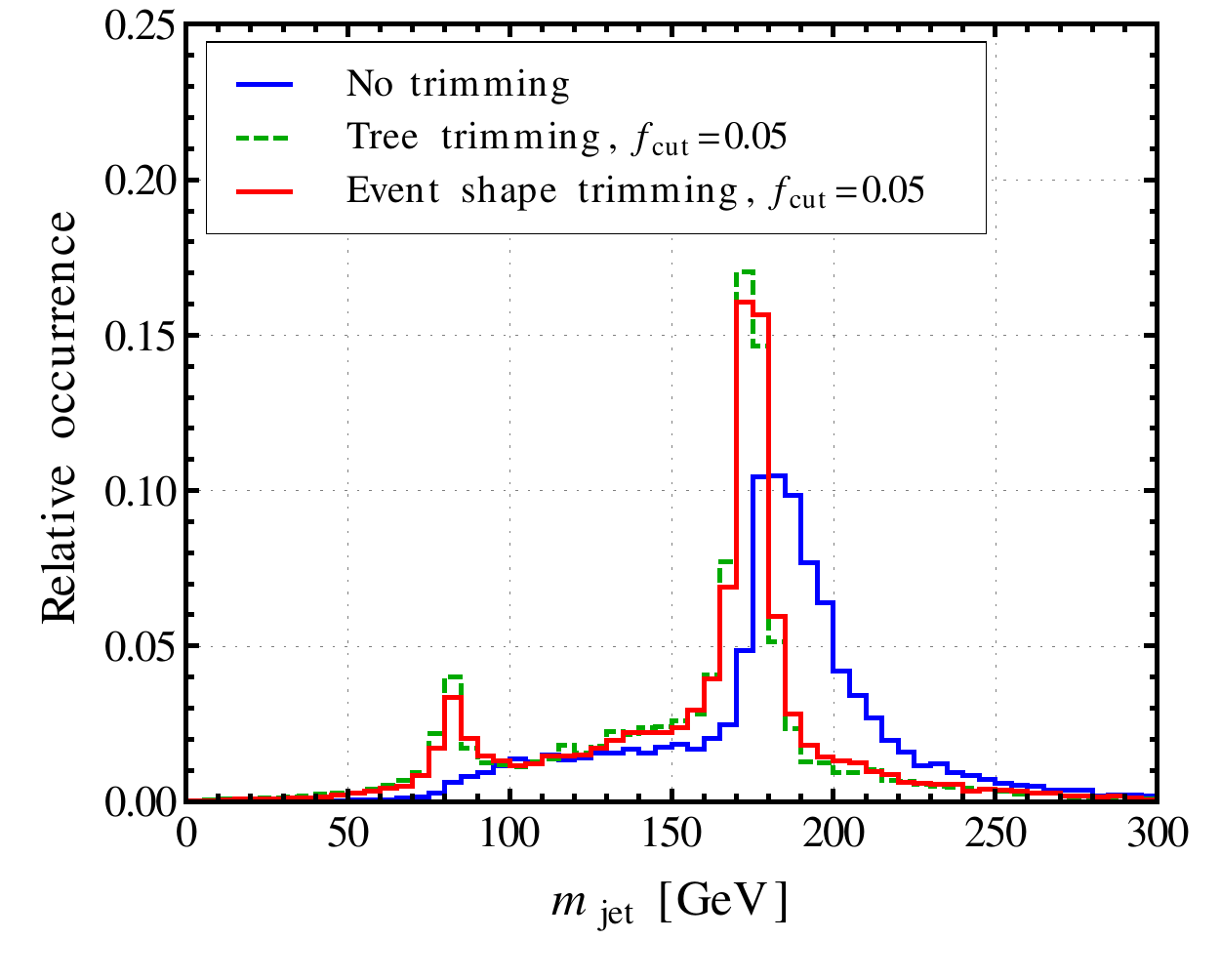}}   
\hspace{0.3in}
  \subfloat[]{\label{fig:TrimEvent2}
    \includegraphics[scale=0.55]{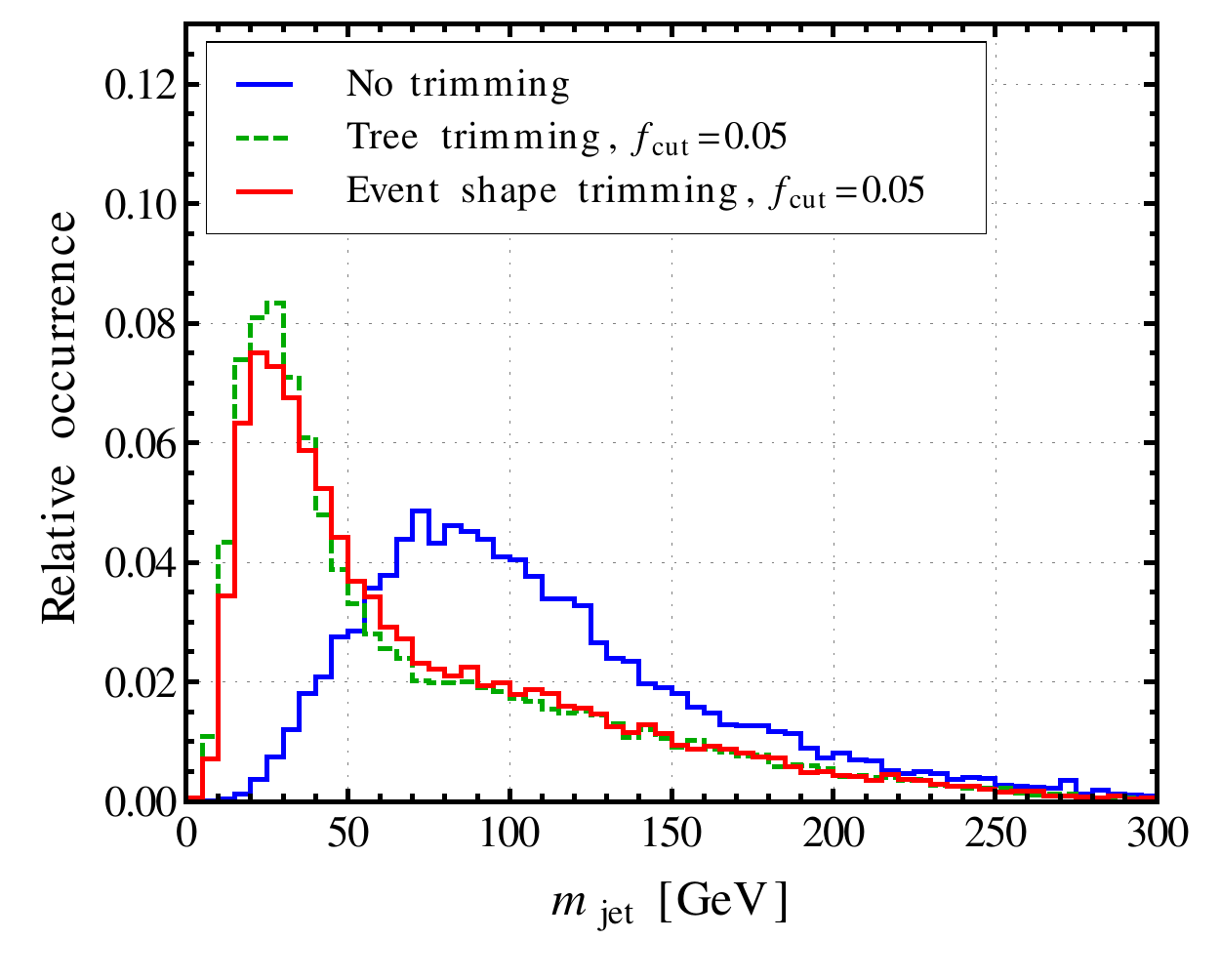}}
  \caption{Boosted top sample (left) and corresponding QCD background (right) from the BOOST 2010 event samples \cite{Abdesselam:2010pt}.  For ordinary tree trimming, we identify jets anti-$k_T$ jets with $R=1.0$ and $\ptc=200~\GeV$, and then applying trimming with $R_\sub=0.3$ and $f_\text{cut}=0.05$.  For shape trimming, we apply event-wide trimming using the same $R_\sub$ and $f_\text{cut}$ parameter before clustering with anti-$k_T$.  In both cases, we plot the masses of the two hardest jets per event.}
  \label{fig:TrimEvent}
 \end{figure}

To compare the behavior of ordinary tree trimming and shape trimming, we use event samples from the BOOST 2010 report \cite{Abdesselam:2010pt}.  In particular, we analyze a boosted top signal and the corresponding QCD background in the $p_T$ bin $500~\GeV<p_T<600~\GeV$.\footnote{Event samples from BOOST 2010 and details about events generation can be found at \url{http://www.lpthe.jussieu.fr/~salam/projects/boost2010-events/herwig65} and \url{http://tev4.phys.washington.edu/TeraScale/boost2010/herwig65}.  These events are for the 7 TeV LHC generated with \textsc{Herwig}~{6.510} \cite{Corcella:2002jc}, with underlying event given by JIMMY \cite{Butterworth:1996zw} with an ATLAS tune \cite{atlasmc}.}  In \Fig{fig:TrimEvent}, we show the effect of trimming on the jet mass spectrum for the boosted top signal and the corresponding QCD background. For tree trimming, we build anti-$k_T$ jets with $R=1.0$ and $\ptc=200~\GeV$ and trim with $R_\sub=0.3$ and $f_\text{cut}=0.05$.  For shape trimming, we use the same set of parameters to trim the entire event according to weights from \Eq{eq:trimweights}, and then build anti-$k_T$ jets with $R=1.0$ and $\ptc=200~\GeV$.  We see that the behavior of both trimming methods is very similar, and that both methods emphasize the boosted top mass peak while suppressing the high-mass QCD background.  Jet shape trimming is not shown in \Fig{fig:TrimEvent} as it performs very similarly to event shape trimming.

\begin{figure}[t]
  \centering
      \includegraphics[scale=0.6]{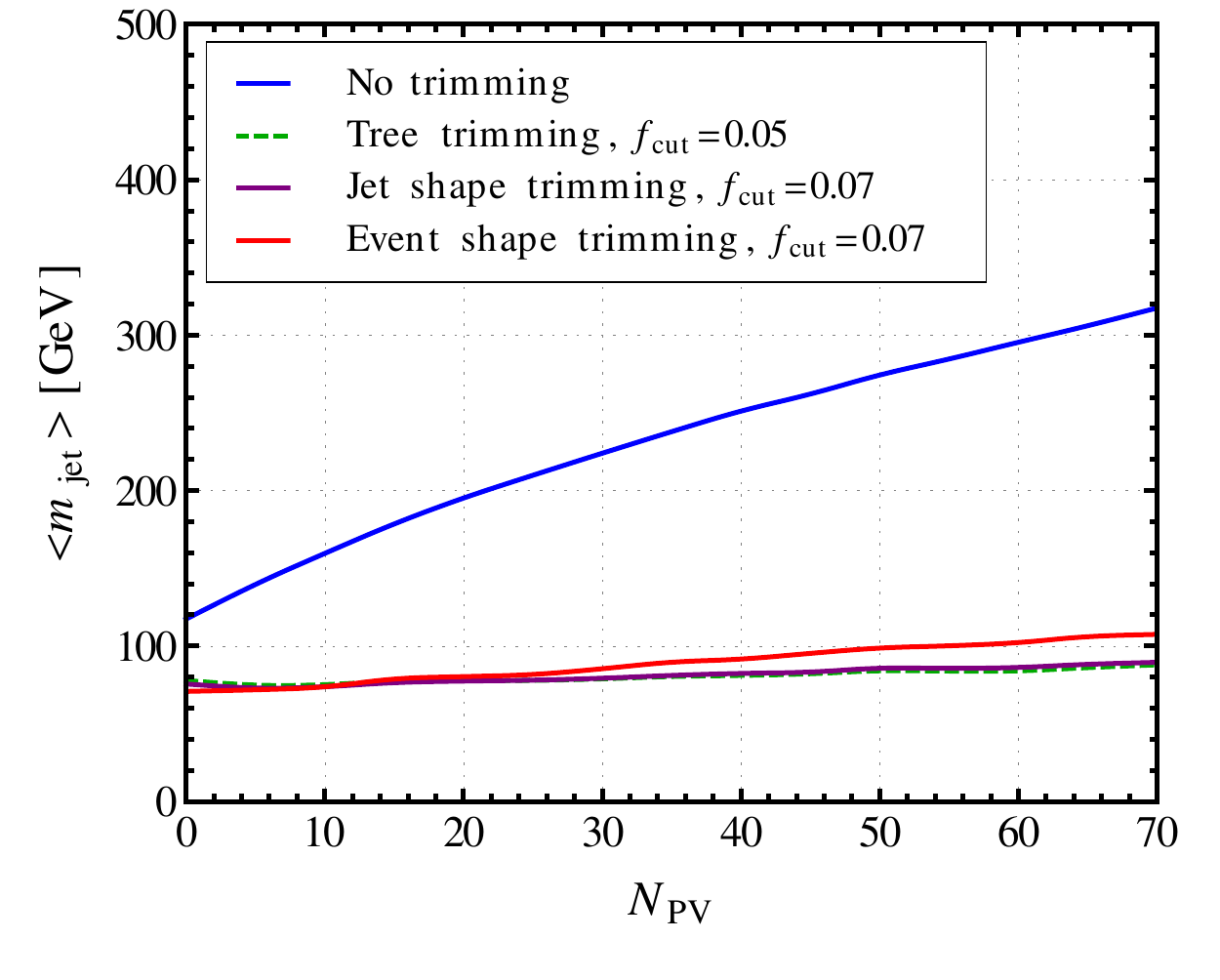}

  \caption{Pileup mitigation for $Z(\rightarrow\nu\nu)+j$ events.  Shown is the mean of the hardest jet mass distribution as a function of the number of primary vertices $N_{\text{PV}}$.  In all cases we use $R_\sub=0.3$, with  $f_\text{cut}=0.05$ for tree-trimming and $f_\text{cut}=0.07$ for the two shape trimming options.
}
  \label{fig:pileup}
 \end{figure}

One important application for trimming is pileup mitigation \cite{Chatrchyan:2013rla,Aad:2013gja}.  To study its effectiveness, we take our sample of $Z(\rightarrow\nu\nu)+j$ events from \Sec{sec:JetObs} and overlay $N_{\text{PV}}$ soft QCD events generated with \pythia~8.157 \cite{Sjostrand:2007gs}.\footnote{Here, the minimum $p_T$ for the hard process at generator level has been reset to $\ptc^{\text{parton}}=350$~GeV.} We consider three options:  ordinary tree trimming, shape trimming applied to the individual jets (jet shape trimming), and shape trimming applied to the entire event (event shape trimming).   \Fig{fig:pileup} shows the average of the hardest jet mass as a function of $N_{\text{PV}}$, where the jets are built using anti-$k_T$ with $R=1.0$ and $\ptc=500~\GeV$.  Taking $R_\sub=0.3$ in all cases, we find a comparable degree of stability against pileup for tree trimming with $f_\text{cut}=0.05$ (as was done in \Ref{Aad:2013gja}), jet shape trimming with $f_\text{cut}=0.07$, and event shape trimming with $f_\text{cut}=0.07$.  Note that event shape trimming has the largest variation with $N_{\text{PV}}$, as expected since $p_{T i,R}$ is typically lower than $p_{T \jet}$, and therefore does not groom as aggressively.  Part of the reason we need a different $f_\text{cut}$ value for tree trimming versus shape trimming is that the effective subjet areas of the two methods are different.

A complementary way to do pileup mitigation is via area subtraction \cite{Cacciari:2007fd,Cacciari:2008gn,Soyez:2012hv}.  It is straightforward to correct the trimming weights in \Eq{eq:trimweights} using area subtraction, because $p_{T i,R_\sub}$ and $p_{T i,R}$ are defined in terms of fixed-radius regions, and therefore have fixed areas $\pi R_\sub^2$ and $\pi R^2$ respectively.  At present, our \FastJet\ implementation of shape trimming does not include area subtraction, but we anticipate including that functionality in a future version.

\section{Generalizations}
\label{sec:Generalizations}

\subsection{Other Jet-like Event Shapes}
\label{subsec:singlejet}

\begin{figure}[p]
  \centering
  \subfloat[]{\label{fig:HistoHT_neg1}
    \includegraphics[scale=0.65]{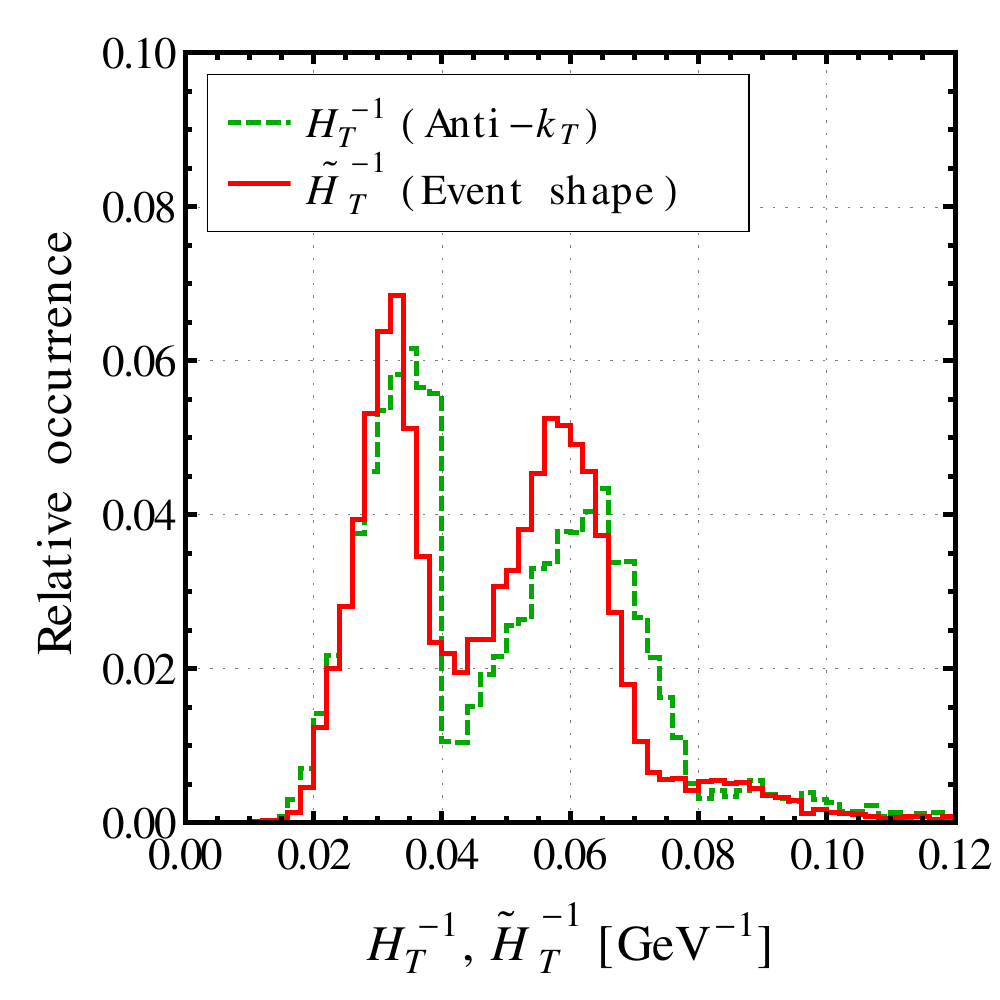}}   
\hspace{0.3in}
  \subfloat[]{\label{fig:ScatterHT_neg1}
    \includegraphics[scale=0.65]{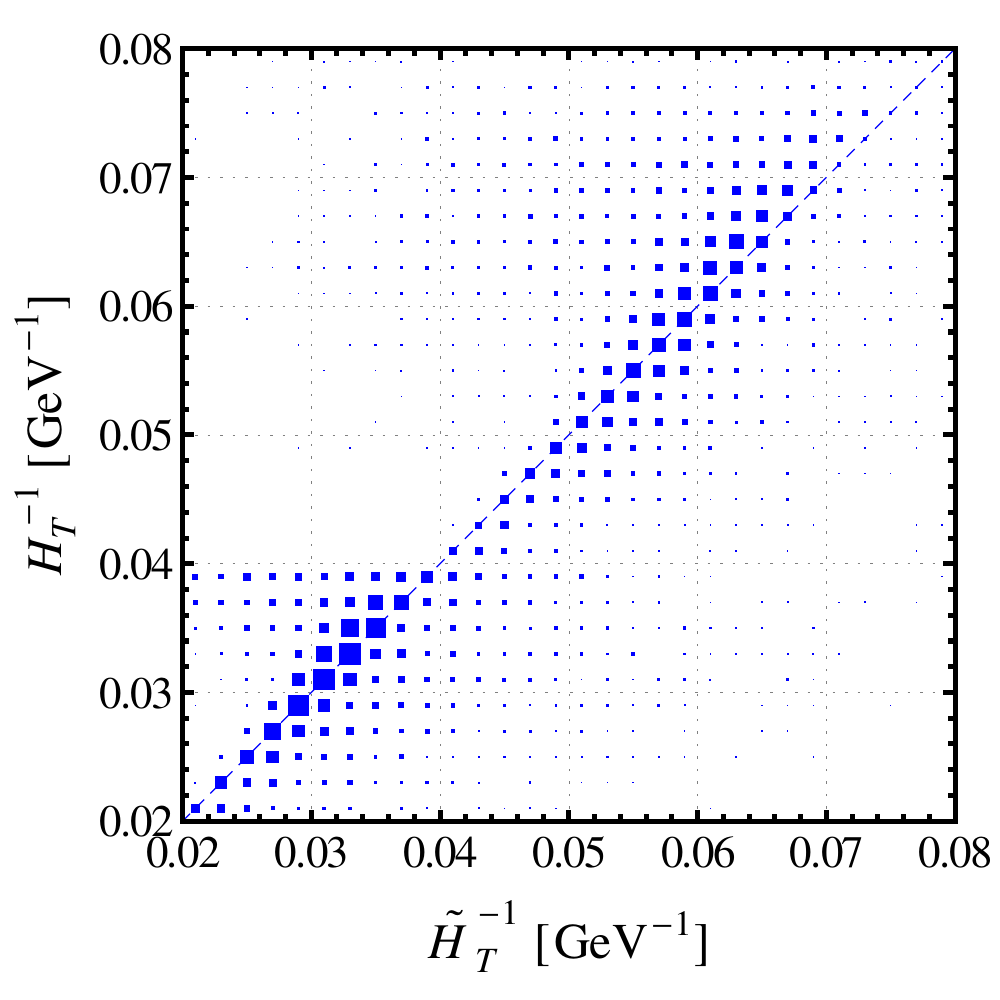}}
  \caption{Summed transverse momentum inverse (i.e.\ $H^n_T$ with $n = -1$) for QCD dijet events.  The jet parameters, formatting, and cuts are the same as for \Fig{fig:Njet}.}
  \label{fig:HT_neg1}
 \end{figure}
 
The general procedure to build event shapes $\widetilde{\F}$ from single jet observables $\F_{\jet}$ was given in \Sec{sec:JetObs}.  Here we give a few more examples beyond $\widetilde{N}_\jet$, $\widetilde{H}_T$, and $\widetilde{\slashed{p}}_T$.

As a simple generalization of $N_\jet$ and $H_T$, consider the jet-based observable
\be
H^n_T(\ptc,R) =\sum_\jets p_{T\jet}^n\, \Theta(p_{T\jet}-\ptc),
\ee
where $n = 0$ ($n=1$) corresponds to $N_\jet$ ($H_T$).  Using the method in \Sec{sec:JetObs}, the corresponding event shape is
\be
\label{eq:GeneralizedHT}
\widetilde{H}^n_T(\ptc,R) = \sum_{i\in\ev} \frac{p_{T,i}}{p_{Ti,R}} (p_{Ti,R})^n \, \Theta(p_{T i,R}-\ptc).
\ee
In \Fig{fig:HT_neg1}, we compare $H^n_T(\ptc,R)$ to $\widetilde{H}^n_T(\ptc,R)$ for $n=-1$ in QCD dijet events, using the same event generation scheme as \Sec{sec:JetObs}.

 \begin{figure}[p]
  \centering
  \subfloat[]{\label{fig:HistoMj}
    \includegraphics[scale=0.65]{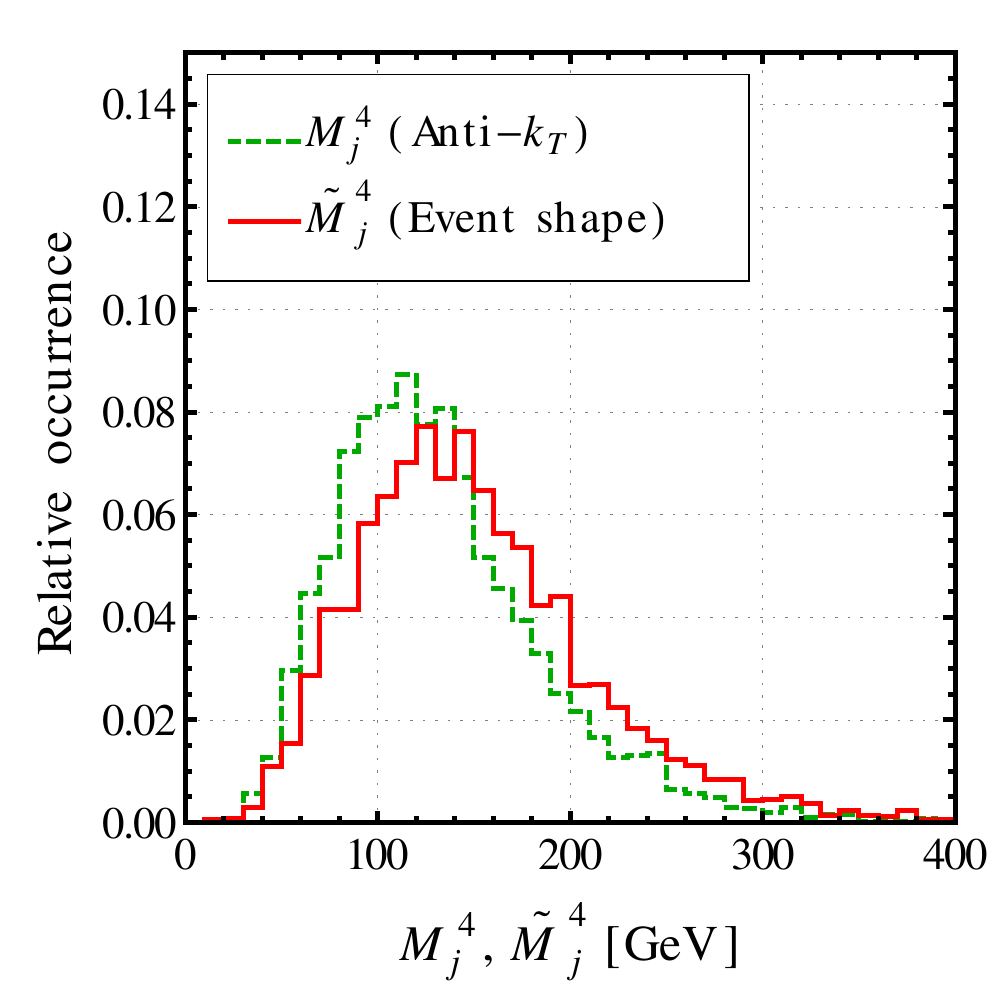}}   
\hspace{0.3in}
  \subfloat[]{\label{fig:ScatterMj}
    \includegraphics[scale=0.65]{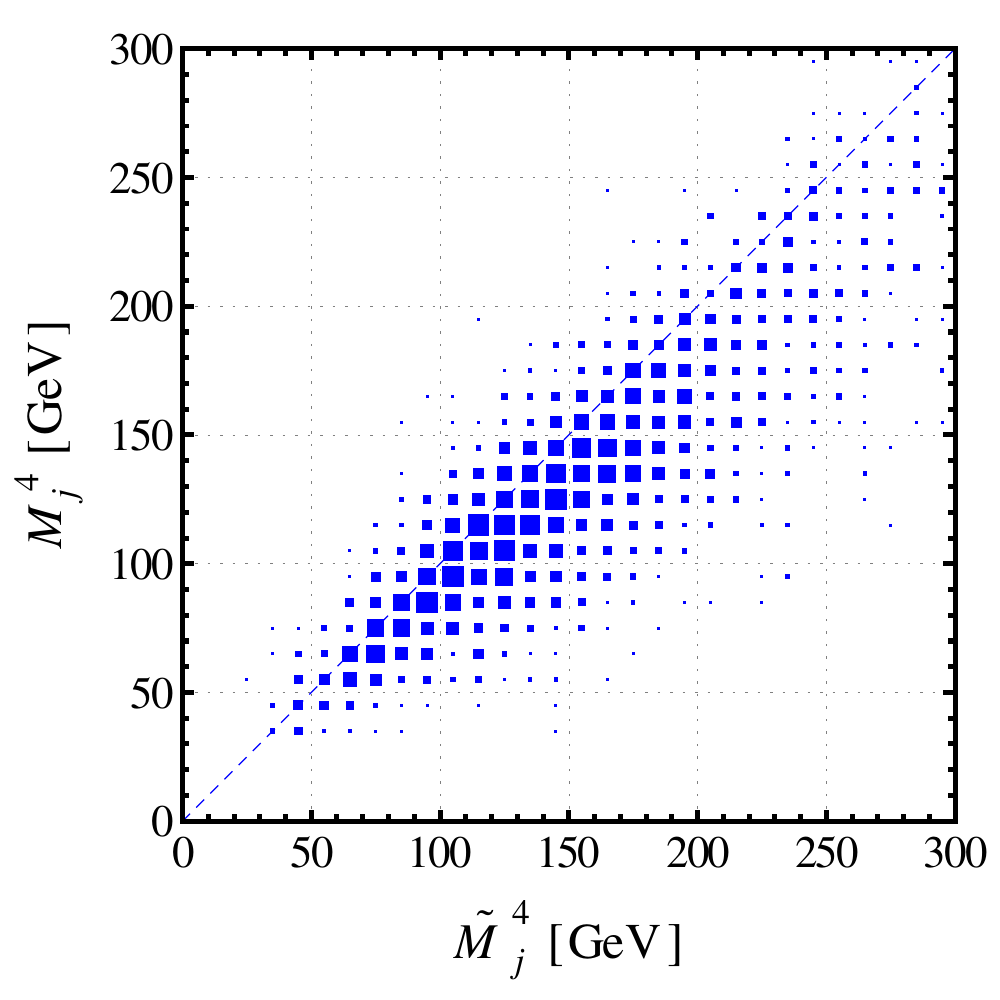}}
  \caption{Summed jet mass analysis for $M_J^4$ that mimics \Ref{Hedri:2013pvl}. Shown is a QCD four-jet sample with the (trimmed) summed jet mass of the four hardest jets.  For the anti-$k_T$ version, the trimmed jets have $R=1.2$, $\ptc = 50~\GeV$, $R_\text{sub}=0.3$, and $f_\text{cut}=0.05$, requiring at least four such jets and the hardest jet above $100~\GeV$.  For the event shape version, the event selection criteria is $\widetilde{N}^\text{trim}_\jet > 3.5$ and $\widetilde{p}^{\,\text{trim}}_T(1) > 100~\GeV$ with the same jet and trimming parameters above, and the observable is $\widetilde{M}^{4}_J$ defined in \Eq{eq:MJtrim4}, calculated after the trimming weights in \Eq{eq:trimweights} are applied.}
  \label{fig:summedMass}
 \end{figure}

A more complicated example is the sum of jet masses in an event,
\be
M_J(\ptc,R) = \sum_\jets m_\jet\, \Theta(p_{T\jet}-\ptc).
\ee
The corresponding event shape is given by
\be
\widetilde{M}_J(\ptc,R) =  \sum_{i\in\ev} \frac{p_{T,i}}{p_{Ti,R}} m_{i,R} \, \Theta(p_{T i,R}-\ptc),
\ee
where $m_{i,R} = \sqrt{|p_{i,R}^\mu|^2}$.  One could of course raise $m_{i,R}$ to a power in analogy with $\widetilde{H}^n_T$.

Summed jet mass is a potentially powerful variable to study high jet multiplicity events at the LHC \cite{Hook:2012fd}, and can be combined with other substructure observables to control QCD multijet backgrounds to new physics searches \cite{Cohen:2012yc,Hedri:2013pvl}.  As an example, it is instructive to see how to mimic aspects of such an analysis using event shapes.  In \Ref{Hedri:2013pvl}, events were clustered into fat jets with $R = 1.2$, the fat jets were trimmed ($R_\text{sub}=0.3$, $f_\text{cut}=0.05$), and events were retained if they had at least four fat jets above $\ptc = 50~\GeV$ and the hardest jet above $100~\GeV$.  Then the (trimmed) summed jet mass was taken for just the four hardest jets.  To mimic the selection procedure, one would take events with $\widetilde{N}^\text{trim}_\jet(\ptc,R;f_\text{cut},R_\text{sub}) > 3.5$ (see \Eq{eq:Ntrim}) and $\widetilde{p}^{\,\text{trim}}_T(1) > 100~\GeV$.  To mimic the observable, one would first apply the shape trimming weights from \Eq{eq:trimweights}, and then define
\be
\label{eq:MJtrim4}
\widetilde{M}^{4}_J(\ptc,R) \equiv  M_J(\ptc',R), \qquad \ptc' = \max\{\ptc,\widetilde{p}^{\,\text{trim}}_T(5)\},
\ee
where $\ptc'$ effectively picks out the four hardest jets (see \Eq{eq:ptcprime}).  In \Fig{fig:summedMass}, we compare the distributions of the (trimmed) summed mass calculated using the two different methods on a QCD four-jet sample.  Despite the somewhat complicated form of the event shape version, there are clear correlations between the methods.  We will discuss the subjet counting aspect of \Ref{Hedri:2013pvl} in \Sec{sec:subjetevent}.\footnote{The event-subjettiness variable of \Ref{Cohen:2012yc} is defined as a geometric mean of $N$-subjettiness ratios \cite{Thaler:2010tr,Thaler:2011gf} measured on individual jets.  To convert that to an event shape, we would first take the logarithm, since that would correspond to a sum over the logs of individual jet observables, and is therefore in the form needed in \Eq{eq:defF}.}

\subsection{Subjet-like Jet Shapes}

Thus far, we have focused on jet-like event shapes, but it is clear that the same technique can be applied to subjet-like jet shapes.  These jet shapes would probe the substructure of a given jet, and can be defined according to \Eqs{eq:defF}{eq:generalForm} with  ``jet'' replaced by ``subjet'' and ``event'' replaced by ``jet''.  Concretely, given a jet found using an ordinary jet algorithm, consider a subjet-based observable built from subjets of radius $R_\sub$ above $\ptsubc$:
\be
\label{eq:defG}
\G(\ptsubc,R_\sub) = \sum_\subjs \G_\subj \,  \Theta(p_{T \sub}-\ptsubc),
\ee
where $\G_\subj \equiv g(\{p^\mu_j\}_{j\in \subj})$ depends on the kinematics of the individual subjet constituents.  The corresponding jet shape would be
\be
\label{eq:defGtilde}
\Gt(\ptc,R_\sub)=\sum_{i\in\jet}\frac{p_{Ti}}{p_{Ti,R_\sub}}\G_{i,R_\sub}\,\Theta(p_{T i,R_\sub}-\ptsubc),
\ee
where $\G_{i,R_\sub} \equiv g(\{p^\mu_j\,\Theta(R_\sub-\Delta R_{ij})\}_{j\in\jet})$.

\begin{figure}[t]
  \centering
  \subfloat[]{\label{fig:histoNsubjTop}
    \includegraphics[scale=0.65]{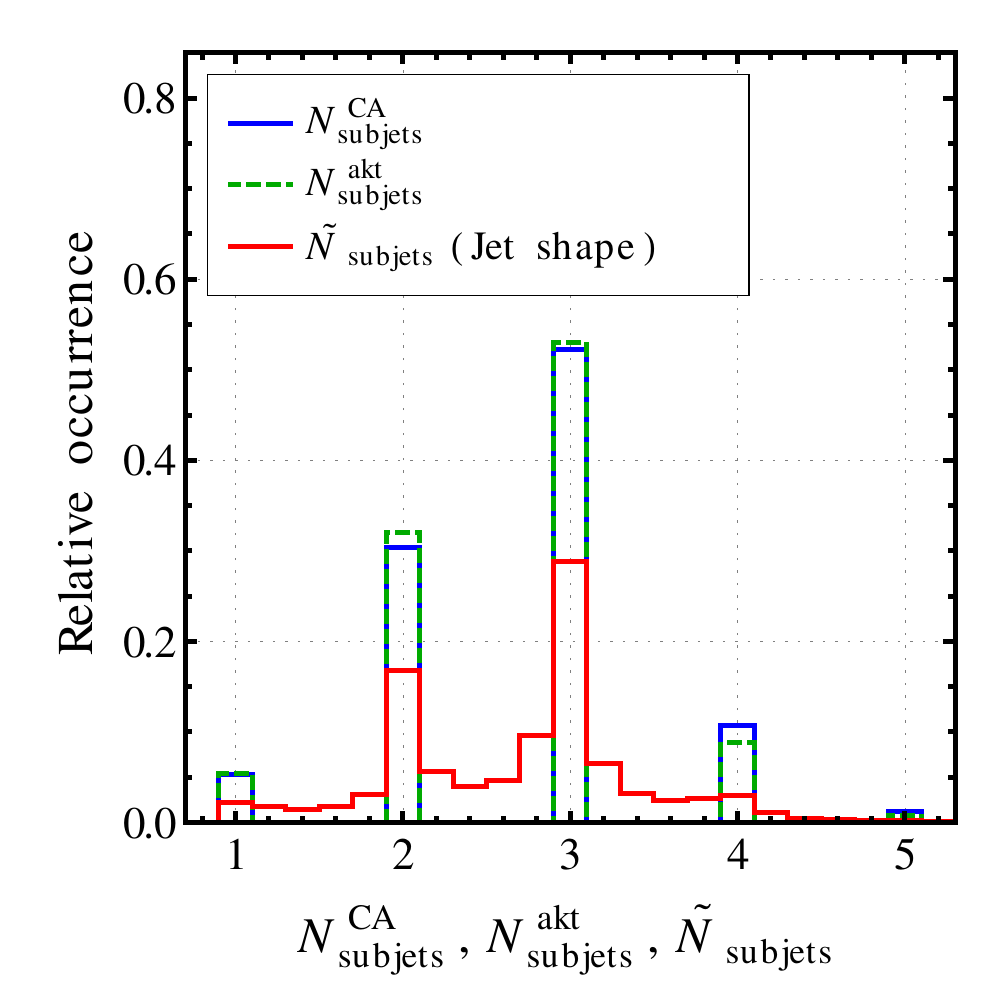}}   
\hspace{0.3in}
  \subfloat[]{\label{fig:scatterNsubjTop}
    \includegraphics[scale=0.65]{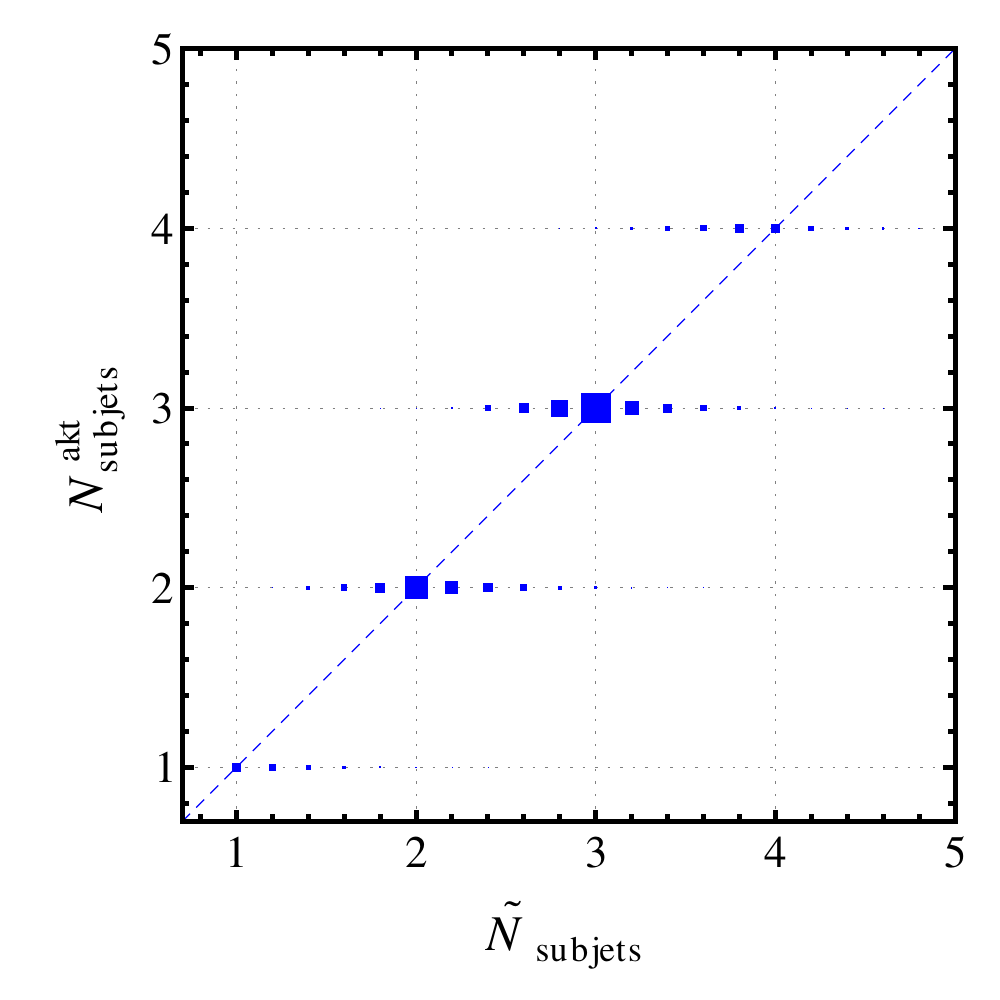}}
  \caption{Subjet multiplicity (i.e.~$N_\subj$) on the boosted top sample from BOOST 2010.  The jet selection is the same as in \Fig{fig:TrimEvent}, and we count subjets either with Cambridge-Aachen clustering, anti-$k_T$ clustering, or $\widetilde{N}_\text{subjet}$.  In all cases, we take $R_\sub=0.3$ and $\ptsubc = 0.05 \, p_{T\jet}$.  In the case of Cambridge-Aachen clustering, this is equivalent to counting the subjets left from (CA) trimming with $R_\sub=0.3$ and $f_\text{cut}=0.05$.}
  \label{fig:NsubjTop}
 \end{figure}

As an example, a jet shape that counts the subjet multiplicity is
\be
\label{eq:Nsub}
\widetilde{N}_\text{subjet}(\ptsubc,R_\sub) = \sum_{i\in \jet} \frac{p_{T i}}{p_{T i,R_\sub}}\Theta(p_{T i,R_\sub}-\ptsubc).
\ee
In \Fig{fig:NsubjTop} we study subjet multiplicity for the same boosted top sample analyzed in \Sec{sec:Trimming}.  Starting from anti-$k_T$ jets with $R=1.0$ and $\ptc=200~\GeV$, we count the number of subjets in three different ways.  First, we count the number of Cambridge-Aachen subjets left after trimming is applied with $R_\sub=0.3$ and $f_\text{cut}=0.05$.  Second, we re-run anti-$k_T$ clustering on the jet with $R_\sub=0.3$ and $\ptsubc = f_\text{cut}\,p_{T\jet}$.  Third, we use the jet shape $\widetilde{N}_\text{subjet}$ with the same value of $R_\sub$ and $\ptsubc$.  The first two methods necessarily yield integer values, whereas $\widetilde{N}_\text{subjet}$ is continuous.  All three methods peak at $N_\text{subjet} = 3$, as expected since this is a boosted top quark sample.

\subsection{Subjet-like Event Shapes}
\label{sec:subjetevent}

Our final generalization is to observables that are inclusive over the subjets in an entire event.  That is, we want to start from an observable defined in terms of the constituents in a subjet, summed over all subjets in each jet, and then further summed over all jets in the event.  Consider an observable built from jets of radius $R$ above $\ptc$ with subjets of radius $R_\sub$ above $\ptsubc$:
\be
\label{eq:defH}
\HH(\ptc,R_\sub; \ptsubc,R_\sub)=\sum_\jets\,\sum_\subjs\, \HH_\subj \Theta(p_{T \sub}-\ptsubc) \Theta(p_{T \jet}-\ptc),
\ee
where $\HH_\subj\equiv h(\{p^\mu_j\}_{j\in \subj})$ depends on the kinematics of the subjet constituents.  The corresponding event shape is
\be
\label{eq:defHtilde}
\widetilde{\HH}(\ptc,R; \ptsubc,R_\sub) = \sum_{i\in\ev} \frac{p_{Ti}}{p_{Ti,R_\sub}} \HH_{i,R_\sub} \,\Theta(p_{T i,R_\sub}-\ptsubc) \Theta(p_{T i,R}-\ptc),
\ee
where $\HH_{i,R_\sub} \equiv h(\{p^\mu_j\,\Theta(R_\sub-\Delta R_{ij})\}_{j\in\ev})$.  Note that the weight factor depends on $p_{Ti,R_\sub}$, and $p_{Ti,R}$ only appears for testing $\ptc$. 

For measurement functions $\HH_\subj$ that are expressible as a sums over the subjet constituents,
\be
\HH_\subj= \sum_{j\in \subj} \tilde{h}(p^\mu_j),
\ee
where $\tilde{h}$ is a single particle measurement function, we can elide the $p_{Ti}/p_{Ti,R_\sub}$ weighting factor and directly write down the event shape
\be
\label{eq:altdefHtilde}
\widetilde{\HH}(\ptc,R; \ptsubc,R_\sub) = \sum_{i\in\ev} \tilde{h}(p^\mu_i) \,\Theta(p_{T i,R_\sub}-\ptsubc) \Theta(p_{T i,R}-\ptc).
\ee
The shape trimming technique from \Sec{sec:Trimming} can be expressed as such an event shape, with $\ptsubc = f_\cut \,p_{Ti,R}$ and $\tilde{h}(p^\mu_j) = p^\mu_j$ (see \Eq{eq:ttev}).

\begin{figure}[t]
  \centering
  \subfloat[]{\label{fig:histoNsubj}
    \includegraphics[scale=0.65]{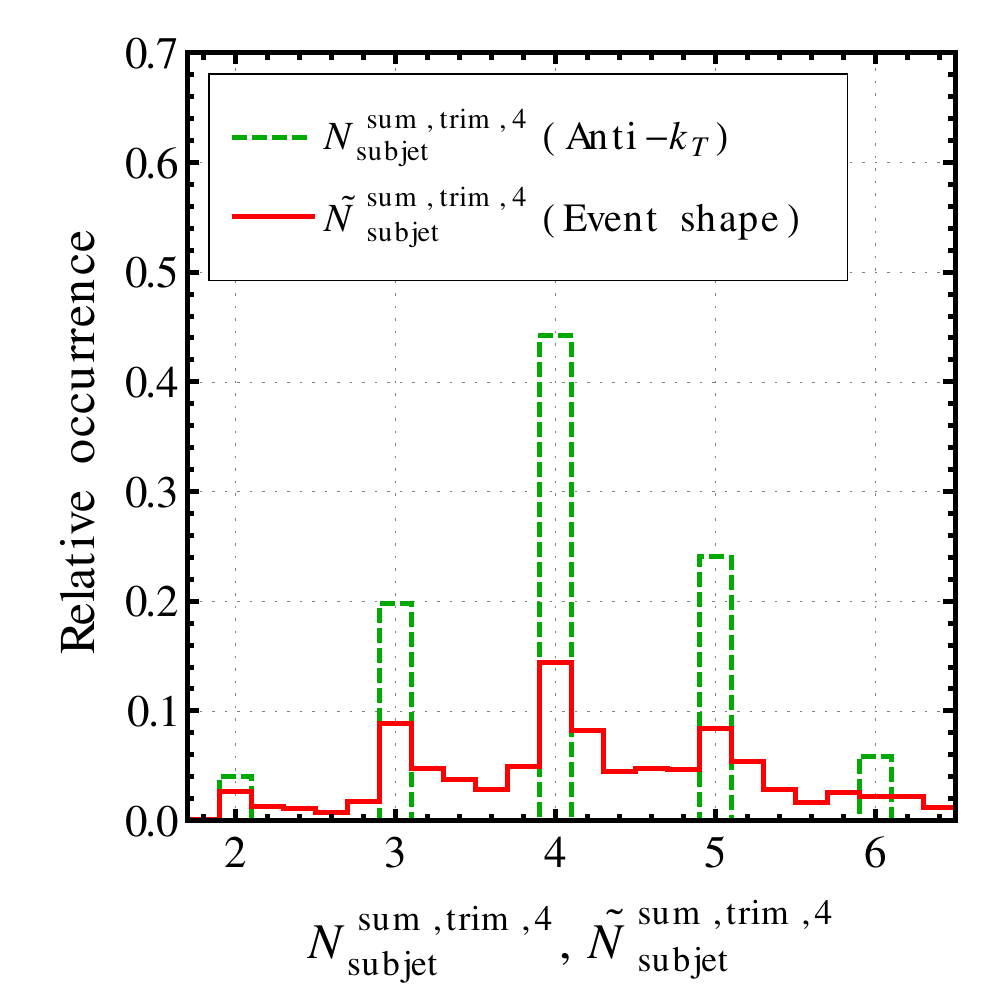}}   
\hspace{0.3in}
  \subfloat[]{\label{fig:scatterNsubj}
    \includegraphics[scale=0.65]{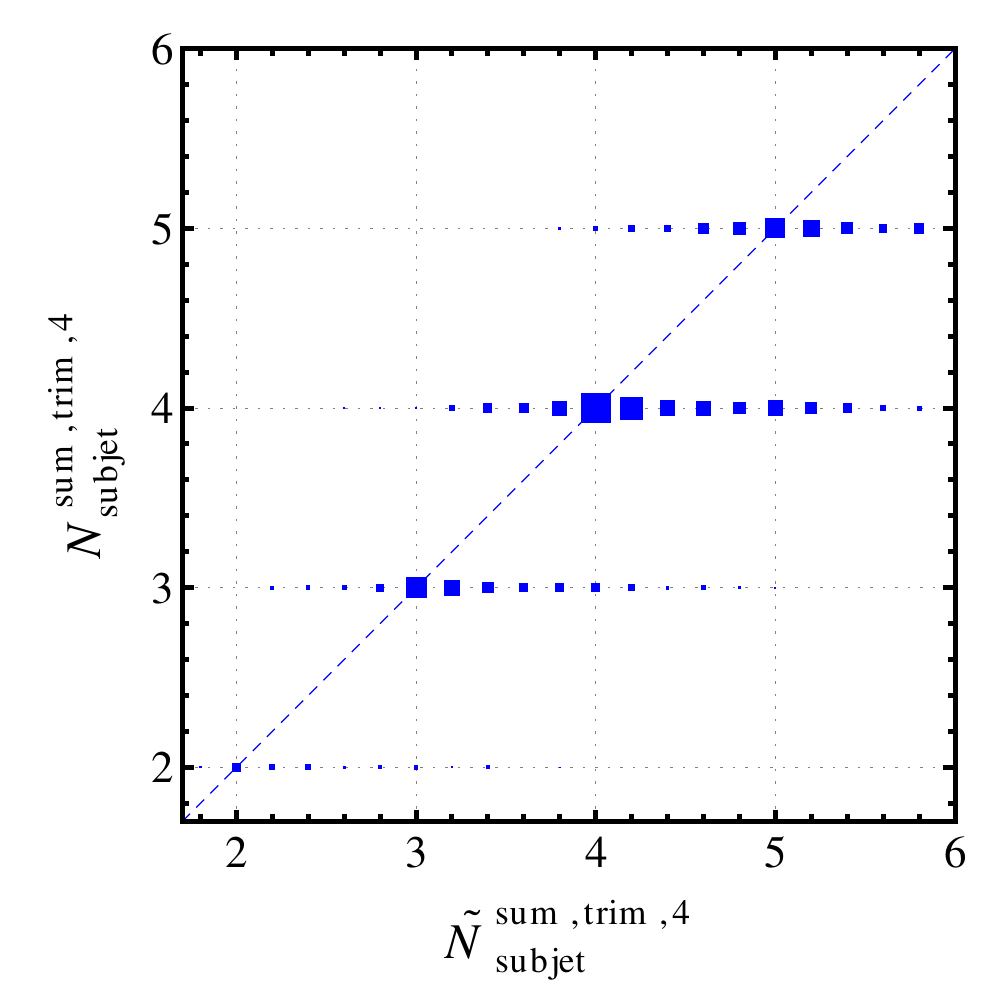}}
  \caption{Summed subjet multiplicity (i.e.~$N^\text{sum}_{\subj}$) on a QCD four-jet sample, in the spirit of \Ref{Hedri:2013pvl}.  The jet selection is the same as \Fig{fig:summedMass}, and we use the event shape $\widetilde{N}^\text{sum,trim,4}_\text{subjet}$ from \Eq{eq:nsub4}.}
  \label{fig:Nsubj}
 \end{figure}

Following the example of subjet multiplicity $\widetilde{N}_\text{subjet}$ in \Eq{eq:Nsub}, we can define the (trimmed) summed subjet multiplicity:
\begin{align}
&\widetilde{N}^{\text{sum},\trim}_{\text{subjet}}(\ptc,R; f_{\rm cut}, \ptsubc, R_\sub) \\
&\qquad ~= \sum_{i\in\ev}  \frac{p_{T i}}{p_{T i,R_\sub}} \, \Theta(p_{T i,R_\sub}-\ptsubc) \Theta(\frac{p_{T i,R_\sub}}{p_{T i,R}}-f_\cut) \Theta(p_{T i,R}-\ptc) \nonumber,
\end{align}
where the trimming criteria on the subjets is only imposed if it is stricter than the $\ptsubc$ requirement.  A similar variable was used in \Ref{Hedri:2013pvl} to isolate high jet multiplicity events at the LHC, in concert with the summed jet mass already mentioned in \Sec{subsec:singlejet}.  Here, however, we are restricted to defining subjets with a fixed radius $R_\sub$, as opposed to the more dynamical subjet finding procedures advocated in \Ref{Hedri:2013pvl}.\footnote{In principle, one could choose the subjet radius $R_\sub$ to be a (local) function of the particles within a radius $R$ of particle $i$.}  In \Fig{fig:Nsubj}, we compare subjet counting using anti-$k_T$ for both fat jets and subjets to the comparable procedure with $\widetilde{N}^\text{sum}_\text{subjet}$ on the QCD four-jet sample.  We use the same event selection as in \Sec{subsec:singlejet}, and define
\be
\label{eq:nsub4}
\widetilde{N}^\text{sum,trim,4}_\text{subjet}(\ptc,R; f_{\rm cut}, \ptsubc, R_\sub) = \widetilde{N}^\text{sum,trim}_\text{subjet}(\ptc',R; f_{\rm cut}, \ptsubc, R_\sub)
\ee
with $\ptc' =  \max\{\ptc,\widetilde{p}^{\,\text{trim}}_T(5)\}$ to effectively isolate the four hardest jets.  Apart from the non-integer nature of $\widetilde{N}^\text{sum,trim,4}_\text{subjet}$, there is a clear correlation between the methods.

\section{Conclusions}
\label{sec:Conclusions}

In this paper, we have shown how inclusive jet observables can be recast as jet-like event shapes.  By replacing an inclusive sum over jets in an event with an inclusive sum over particles in an event, we have removed the dependence on the jet clustering procedure, while still maintaining the jet-like radius $R$ and jet-like momentum cut $\ptc$ expected in jet-based analyses.  While our original method can only be applied to inclusive jet observables, we have shown one example where more exclusive information about single jets was obtained by inverting the jet multiplicity event shape $\widetilde{N}_\jet$ to determine the $p_T$ of the $n$-th hardest jet.  Our focus was on event shapes in this paper, though we have shown that there is a straightforward generalization to jet shapes, which may find use in jet substructure studies.

A promising possible application of these event shapes is for event selection at the trigger level, especially given their local computational structure.  To the best of our knowledge, all jet triggers presently in use on the ATLAS and CMS experiments can be mimicked by appropriate combinations of  $\widetilde{N}_\jet$, $\widetilde{H}_T$, and $\widetilde{\slashed{p}}_T$ cuts (choosing different values of $R$ and $\ptc$ as needed).  It may even be possible to do preliminary jet identification at the trigger level using the hybrid event shapes with winner-take-all recombination; the local nature of the clustering means that the approach can be parallelized across the detector without double-counting.  Of course, more detailed feasibility studies are needed to see whether these event shapes can be incorporated into the trigger upgrades planned for high-luminosity LHC running.

For analysis-level jet studies, the event shapes provide a complementary characterization of the gross jet-like nature of the event.  From the correlations seen in \Sec{sec:JetObs}, one should expect $\mathcal{F}$ and $\widetilde{\mathcal{F}}$ to have similar performance in an experimental context.  There can be important differences, however, in regions of phase space where jets are overlapping or otherwise ambiguous.  Thus, a comparison between, say, a selection criteria based on ${N}_\jet$ and one based on $\widetilde{N}_\jet$ would offer a useful test for the robustness of an analysis.

A novel application of our method is for jet grooming via shape trimming.  This worked because ordinary tree trimming \cite{Krohn:2009th} can be written as a double sum over subjets and jets in an event, allowing an application of the general techniques in \Sec{sec:subjetevent}.  Shape trimming can be applied to event shapes themselves, or it can be interpreted as simply assigning a weight to each particle in an event, after which one can perform a traditional jet-based analysis.  Shape trimming has similar pileup mitigation performance to tree trimming, but can be more easily applied event-wide since it does not require the explicit identification of jets or subjets.

Other grooming techniques beyond trimming deserve future study, though we do not know (yet) how to cast them as event shapes.  For example, filtering \cite{Butterworth:2008iy} is based on keeping a fixed number of subjets, which we do not know how to implement as an inclusive sum over all particles in an event.  Similarly, pruning \cite{Ellis:2009su,Ellis:2009me} and (modified) mass drop \cite{Butterworth:2008iy,Dasgupta:2013ihk,Dasgupta:2013via} are based on recursively applying a selection criteria, which have no obvious event shape counterpart.  The modified mass drop procedure is particularly interesting because it removes Sudakov double logarithms \cite{Dasgupta:2013ihk,Dasgupta:2013via}, and a non-recursive event shape version of this procedure would help for understanding this unique behavior.

Finally, these event shapes are particularly interesting for future analytic studies in perturbative QCD.  Formally, an inclusive jet observable $\mathcal{F}$ and its event shape counterpart $\widetilde{\mathcal{F}}$ are exactly equivalent for infinitely narrow jets separated by more than $R$, such that they share the same soft-collinear structure.  Therefore, up to non-singular and power-suppressed terms, we expect $\mathcal{F}$ and $\widetilde{\mathcal{F}}$ to have similar (if not identical) factorization and resummation properties.  That said, there is clearly a difference between the integer-valued jet multiplicity $N_{\rm jet}$ and the continuous event shape $\widetilde{N}_{\rm jet}$, though the difference does not show up until $\mathcal{O}(\alpha_s)$ (for jets separated by more than $R$ but less than $2R$) or $\mathcal{O}(\alpha_s^2)$ (for jets separated by more than $2R$).  We expect that understanding the origin of non-integer $\widetilde{N}_{\rm jet}$ values is likely to shed considerable light on the jet-like nature of QCD.

\acknowledgments{This paper is dedicated to the memory of Tucker Chan, MIT class of 2012.  We thank David Krohn and Steve Ellis for their input in the early stages of this work.  We benefitted from conversations with Philip Harris, Andrew Larkoski, Lucia Masetti, Duff Neill, Gavin Salam, Iain Stewart, David Strom, and Nhan Tran.  We thank the \FastJet\ authors (Matteo Cacciari, Gavin Salam, and Gregory Soyez) for helpful feedback on our code.  D.B.\ thanks Andrea Allais for helpful discussions, and J.T.\ thanks Joe Virzi for conversations in 2008 about the possibility of a continuous jet counting observable.  This work is supported by the U.S. Department of Energy (DOE) under cooperative research agreement DE-FG02-05ER-41360.  D.B. is partly supported by the LHC-Theory Initiative Graduate Fellowship and by Istituto Nazionale di Fisica Nucleare (INFN) through the Bruno Rossi Fellowship.  J.T.~is supported by the DOE Early Career research program DE-FG02-11ER-41741 and by a Sloan Research Fellowship from the Alfred P.\ Sloan Foundation.}

\appendix

\section{Inverting Jet Multiplicity}
\label{app:invert}

In \Sec{sec:indivjet}, we want to find the pseudo-inverse of $\widetilde{N}_\jet(\ptc,R)$ as a function of $\ptc$ to get the function $\widetilde{p}_T(n,R)$.  Here, we provide a computationally efficient way to perform this inverse.  Consider a general function of the form
\be
f(c) = \sum_{i=1}^N f_i \, \Theta(c_i - c),
\ee
where $f_i$ and $c_i$ are properties of the $i$-th particle, and $c$ is some value of a cut.  We wish to calculate the pseudo-inverse $c(f)$, which exists because $f(c)$ is a monotonically decreasing function of $c$.  There is an ambiguity in the inverse because $f(c)$ is a step-wise function (with $N$ steps), so there exists a range of values for $c$ with the same value of $f$.

First, construct a list of length $N$ with all of the values of $c_i$, keeping track of the corresponding value of $f_i$ for each entry:  
\be
\{ \{c_1,f_1\}, \{c_2,f_2\}, \ldots, \{c_N, f_N\} \}.
\ee
This list can be sorted from highest value of $c_i$ to lowest value of $c_i$ with computational scaling $N \log N$.  Let $i_j$ be the particle number $i$ for the $j$-th highest element in the sorted list:
\be
\{ \{c_{i_1},f_{i_1}\}, \{c_{i_2},f_{i_2}\}, \ldots, \{c_{i_N}, f_{i_N}\} \},  
\ee
with $c_{i_1} > c_{i_2} > \ldots > c_{i_N}$.  (If there are two value of $c_i$ that are truly identical, one can add a small offset to arbitrarily break the degeneracy.)

From the sorted list, one then calculates the cumulative totals for the corresponding $f_{i_j}$:
\be
\label{eq:fc}
\{ \{c_{i_1},f_{i_1}\}, \{c_{i_2},f_{i_1} + f_{i_2}\}, \ldots, \{c_{i_N}, \sum_{j = 1}^N f_{i_j}\} \}.
\ee
\Eq{eq:fc} gives the function $f(c)$. To find the pseudo-inverse $c(f)$, one finds the value of $J$ such that
\be
\sum_{j = 1}^J f_{i_j} < f < \sum_{j = 1}^{J+1} f_{i_j}.
\ee
(For a sorted list, the computational cost of searching scales like $\log N$.)  The pseudo-inverse $c(f)$ can then take on any value between $c_{i_J}$ and $c_{i_{J+1}}$.  For concreteness, we use
\be
c(f) = c_{i_{J+1}},
\ee
which makes sure that calculating $\widetilde{p}_T(n ,R)$ with $0 < n_{\rm off} < 1$ on infinitely narrow jets gives back the $p_T$ of the $n$-th jet.

\bibliographystyle{JHEP}

\bibliography{JetsWithoutJets}

\end{document}